# Multiscale modelling of the morphology and spatial distribution of θ' precipitates in Al-Cu alloys

H. Liu[a], B. Bellón[a], J. LLorca[a,b,1]

[a] IMDEA Materials Institute, C/Eric Kandel 2, Getafe 28906 – Madrid, Spain
[b] Department of Materials Science. Polytechnic University of Madrid. E. T. S. de Ingenieros de Caminos. 28040 – Madrid, Spain

**Abstract**

A multiscale approach based on the phase-field model is developed to simulate homogeneous and heterogeneous formation of θ' precipitates during high temperature ageing in Al-Cu alloys. The model parameters that determine the different energy contributions (chemical free energy, interfacial energy, lattice parameters, elastic constants) were obtained from either computational thermodynamics databases or from first-principles density functional theory and molecular statics simulations. From the information, the evolution and equilibrium morphology of the θ' precipitates is simulated in 3D using the phase-field model. The model was able to reproduce the evolution of the different orientation variants of plate-like shaped θ' precipitates with orientation relationship $(001)_{\theta'}//(001)_\alpha$ and $[100]_{\theta'}//[100]_\alpha$ during homogeneous nucleation as well as the heterogeneous nucleation on dislocations, leading to the formation of precipitate arrays. Heterogeneous nucleation on pre-existing dislocation(s) was triggered by the interaction energy between the dislocation stress field and the stress-free transformation strain associated to the nucleation of the θ' precipitates. Moreover, the mechanisms controlling the evolution of the morphology and the equilibrium aspect ratio of the precipitates were ascertained. All the predictions of the multiscale model were in good agreement with experimental data.

---

[1] Corresponding Author.
Email address: javier.llorca@imdea.org (J. LLorca)

# 1. Introduction

   Precipitation hardening is well-established as one of the most efficient strategies to increase the yield strength of metallic alloys [1–3]. Precipitates are normally intermetallic particles with sizes in the range from a few to a few hundred nm which appear during ageing. They hinder the glide of dislocations that have to by-pass or shear the precipitates, increasing the critical resolved shear stress to move the dislocation in the slip plane. The strengthening effect of the precipitates depends on a number of factors, which include their size, shape [4–7] and spatial distribution [8–11]. Precipitation hardened alloys are usually subjected after casting to a homogenization treatment above the *solvus* temperature followed by quenching, which leads to a supersatured solid solution of the solute atoms. Afterwards, precipitation is promoted by ageing the alloy at intermediate temperatures (sometimes in combination with mechanical deformation) and the final precipitate structure can be controlled up to some extent from the ageing temperature and time [2–3, 12]. In general, it is accepted that the highest hardening is provided by uniform distributions of precipitates with large aspect ratio but the optimum combination of precipitate size, shape and spatial distribution depends on many factors, including the actual number of slip systems, the critical resolved shear stress of each system, the presence of other deformation mechanisms (such as twinning), etc. Thus, the design of novel precipitation hardened alloys and the optimization of the current ones is based on the ability to determine the precipitate features as a function of the alloy composition and of the precipitation process. However, this task has been carried out so far by means of costly experimental trial and error approaches [13].

   Modelling of precipitation process in engineering is, thus, a very important and complex issue because precipitation is controlled by a number of phenomena that include the chemical free energies and elastic strain energies of the different phases, the interfacial energies between the precipitates and the matrix, the nucleation sites, etc. The classical nucleation and growth theories [14–15] stated the basic mechanisms controlling the formation and growth of precipitates, and remain unchallenged in most aspects, but they could only provide rough qualitative estimations. The phase-field methods built upon these theories and the diffuse interface treatment and gradient thermodynamics of Cahn and Hilliard [16] have emerged as a powerful tool to provide quantitative information the evolution of precipitates during ageing [17–18] as well as the interaction between precipitates and lattice defects, such as dislocations [19].

   From the predictive viewpoint, one of the main challenges of the phase-field modelling strategy is that the input parameters are often empirical or difficult to establish. This includes, for instance, the quantitative thermodynamic description of metastable precipitates, the anisotropic interfacial energy between the matrix and the precipitate, the elastic constants of the precipitate, etc. In order to overcome these limitations, a multiscale modelling strategy based on the phase-field modelling of precipitation [20–21] is presented in this paper and applied to predict the shape and spatial distribution of θ' precipitates in an Al-Cu alloy. This alloy was selected to demonstrate the capabilities of the multiscale approach because it is the one of the alloy systems in which precipitation hardening is more effective [1–2] and often used as an example of precipitation hardened alloys in text books [22]. The parameters of the phase field model were obtained from atomistic simulations performed by means of first-principles Density Functional Theory (DFT) or Molecular Statics (MS) or from well-established thermodynamic descriptions of the system, leading to parameter-free predictions of the precipitate shape and spatial distribution, which were compared with experimental data.

   The characteristics of θ' precipitates in Al-Cu alloys and the construction of the multi-scale model are introduced in sections 2 and 3, respectively. The processing and characterization of the Al-Cu alloys are briefly presented in section 4 and the simulation results are shown and discussed in combination with the experimental observations in section 5. Final conclusions are summarized in section 6.

## 2. Characteristics of θ' precipitates in Al-Cu alloys

The precipitation of binary Al-Cu alloys has been extensively studied from both the experimental and theoretical viewpoints [3, 12, 21, 23–24]. The precipitation sequence in the Al-Cu alloys follows the path [3]:

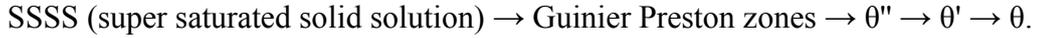

SSSS (super saturated solid solution) → Guinier Preston zones → θ'' → θ' → θ.

The θ' phase is suggested to be the key strengthening phase of this alloy, and its structure is well known [1, 21, 25–26]. θ' is a stoichiometric phase with chemical composition $Al_2Cu$ and tetragonal structure (space group *I4/mmm*, $a = 0.404$ nm, $c = 0580$ nm). The unit cells of α-Al ($Fm\overline{3}m$, $a = 0.404$ nm) matrix and θ' are shown in Figs. 1a and b respectively. The arrangement of the atoms in the $(001)_{θ'}$ plane is similar to $\{001\}_α$ except for the absence of an atom in the face centre, whereas the $(100)_{θ'}$ and $(010)_{θ'}$ planes are quite different from those of α-Al matrix [3]. The orientation relationship between θ' and α-Al is $(001)_{θ'}//(001)_α$ and $[100]_{θ'}//[100]_α$, and the θ' precipitates have three orientation variants. Previous studies [3, 27, 28] revealed the presence of a shear strain when the α-Al lattice transforms to that of θ', and a detailed analysis of the transformation was shown by Nie et al. [3, 28]. It is composed of three successive steps (Figs. 1c-f). The first one is the shift the Al atoms in layers 2 and 3 in opposite directions by a distance $a/6$ (from Fig. 1c to 1d). This is followed by a homogeneous shear of the whole cell by an angle *arctan*(1/3) (from Fig. 1d to 1e) and finally by the shuffle of one Cu atom to the centre of the cell and diffusion of the other Cu atom away to the matrix (Fig. 1f).

According to this phase transformation mechanism, the lattice correspondence between α-Al and θ' is: $[013]_α → [001]_{θ'}$ and $[010]_α → [010]_{θ'}$ [3, 28]. The θ' precipitates were reported to have a plate shape with $\{001\}_α$ habit planes. The broad faces of the θ' plates are nearly fully coherent with the α-Al matrix while the edges of the plates are semi-coherent. It is known that the nucleation of θ' is favoured by the presence of dislocations or pre-existing particles [3] and it has also been reported that θ' precipitates are often randomly distributed in the matrix, but aligned precipitate structures (parallel to the habit plane) can also be observed [2, 3].

Previous quantitative simulations of the equilibrium shape and spatial distribution of θ' precipitates based on the phase-field method were carried out in two dimensions [24, 29]. In addition, the shear strain associated with the α-Al → θ' transformation was not considered. These limitations were overcome in the present investigation, which also took into account the heterogeneous nucleation and growth of θ' precipitates on dislocations.

## 3. Multiscale modelling strategy

The simulation of precipitation is carried out using the meso-scaled phase-field model. The microstructure is described in this context by a set of conserved and non-conserved order parameters, namely $c$ and $\{\eta_p\}$. The former represents the Cu concentration field, i.e., the mole fraction of Cu, while the later stands for the *p*th variants of θ' precipitate. Thus, a position **r** is occupied by the α-Al matrix if $\{\eta_1 = \eta_2 = \cdots = \eta_p = 0\}$ while this position belongs to the *p*th variant of θ' when $\{\eta_p = 1, \text{others} = 0\}$.

The microstructure evolution is expressed by the variation of $c$ and $\{\eta_p\}$ fields with time, which is governed by the Cahn-Hilliard and Allen-Cahn equations [16, 30]:

$$\frac{\partial c}{\partial t} = \nabla \cdot \left[ M \nabla \left( \frac{\delta F}{\delta c} \right) \right] + \xi_c(\mathbf{r}, t), \tag{1}$$

$$\frac{\partial \eta_p}{\partial t} = -L \frac{\delta F}{\delta \eta_p} + \xi_\eta(\mathbf{r}, t), \tag{2}$$

where $M$ is the chemical mobility and $L$ the α-Al/θ' interface mobility. All parameters used in the phase field model equations are dimensionless. The scaling of the energies and the lengths from the dimensionless units to the dimensional units can be found in [31]. In the present work, $M$ was set to 1 and $L$ was set to 5. Thus, the growth of θ' was assumed to be isotropic and the time $t$ stands for a non-dimensional parameter which indicates the number of time steps in the simulation. $\xi_c(\mathbf{r}, t)$ and $\xi_\eta(\mathbf{r}, t)$ stand for the Langevin noise terms for the concentration and the structural order parameters, respectively, which were obtained according to the fluctuation-dissipation theorem [31–34], and are expressed as

$$\xi_c(\mathbf{r}, t) = A_c \nabla \boldsymbol{\rho} \quad \text{and} \quad \xi_\eta(\mathbf{r}, t) = A_\eta \rho, \tag{3}$$

where $\boldsymbol{\rho}$ is a vector in which each component is random number following a Gaussian distribution and $A_c$ and $A_\eta$ stand for the amplitudes of the Langevin noise of the conserved and non-conserved field parameters. According to [31], $A_c = \sqrt{2k_B T M/(v l_0^3 \Delta t)}$ and $A_\eta = \sqrt{2k_B T L/(v l_0^3 \Delta t)}$, where $l_0$ is the grid size, $\Delta t$ time step and $v$ is a scaling factor. More details about the forms of $\xi_c(\mathbf{r}, t)$ and $\xi_c(\mathbf{r}, t)$ can be found in [31–34]. In the case of heterogeneous nucleation of θ' precipitates on dislocations, the Langevin noise strength was chosen in such a way that only heterogeneous nucleation is observed in our simulations. The term $F$ is the total free energy of the system, which is expressed as:

$$F = \int d\mathbf{r} \left[ f(c, \{\eta_p\}) + \frac{\kappa_c}{2} (\nabla c)^2 + \frac{1}{2} \sum_{i=1}^{3} \sum_{j=1}^{3} \sum_{k=1}^{p} \beta_{ij}(p) \nabla_i \eta_k \nabla_j \eta_k \right] + E^{elas} + E^{int}, \tag{4}$$

where $f$ is the chemical free energy, $\kappa_c$ and $\beta_{ij}$ stand for the gradient energy coefficients for the concentration and structural order parameters, respectively. $E^{elas}$ is the elastic strain energy and $E^{int}$ is the interaction energy associated with the formation of θ' phase due to the presence of external stress fields, such as pre-existing dislocations or applied stresses. A detailed account of each energy term is given below.

*3.1 Chemical free energy*

The total chemical free energy, $f(c, \{\eta_v\})$, is constructed on the chemical free energies of the α-Al matrix, $f_\alpha$, and the θ' precipitate, $f_{\theta'}$. The chemical free energy of α-Al matrix is available from the CALPHAD database, and can be described as [35]:

$$f_\alpha = \sum_{i=\text{Al,Cu}} x_i G_i + RT \sum_{i=\text{Al,Cu}} x_i \ln x_i + {}^E G, \tag{5}$$

where $x_i$ is the mole concentration of Al or Cu when $i$ represents Al or Cu. $G_i$ is the molar Gibbs free energy of pure element $i$ and ${}^E G$ is the excess Gibbs free energy, which is expressed by the Redlich-Kister polynomial [36]:

$$G_E = x_{Al}x_{Cu} \sum_{j=0,1} {}^j L_{AlCu}(x_{Al} - x_{Cu})^j. \tag{6}$$

The thermodynamic parameters in Eqs. (4) and (5) are given by [35]

$$G_{Al}^{fcc} = -7976.15 + 137.093038T - 24.3671976T \ln T - 18.84662 \times 10^{-4}T^2 + 74092T^{-1}$$
$$-8.77664 \times 10^{-7}T^3 \quad (298 < T(K) < 700)$$

$$G_{Cu}^{fcc} = -7770.458 + 130.485235T - 24.112392T \ln T - 26.5684 \times 10^{-4}T^2 52478T^{-1}$$
$$-1.29223 \times 10^{-7}T^3 \quad (298 < T(K) < 1358)$$

$${}^0L_{AlCu} = -53520 + 2T \qquad {}^1L_{AlCu} = 38590 - 2T \qquad {}^2L_{AlCu} = 1170$$

and an analytical form of $f_\alpha$ can be derived from Eqs. (4) and (5). For computational efficiency, the variation of the chemical free energy of the α-Al phase as a function of Cu concentration at $T = 200$ °C is approximated by a fourth order polynomial, which is expressed in non-dimensional form as

$$f_\alpha(c) = -1.4632 - 2.9571c - 3.9656c^2 + 5.8588c^3 + 0.8350c^4, \tag{7}$$

Since θ' is a metastable phase, its chemical free energy function is not available. The free energy of θ', $f_{\theta'}$, was approximated by a parabolic function of the solute concentration. A tangent line could then be drawn from the $f_\alpha$ curve at the equilibrium concentration of Cu in the α-Al matrix, which is obtained from the Al-Cu phase diagram [37], to the equilibrium concentration point of θ', as shown in Fig. 2. The equilibrium composition of θ' is considered to be Al$_2$Cu. The only degree of freedom of the parabolic curve is its curvature. The following parabolic function was obtained:

$$f_{\theta'}(c) = -2.9536 - 7.386c + 10c^2, \tag{8}$$

The complete form of the bulk chemical free energy density is expressed as a function of $c$ and $\{\eta_p\}$ as

$$f(c, \{\eta_v\}) = \left[ f_\alpha \left(1 - \sum_{i=1}^p H(\eta_i)\right) + f_{\theta'} \sum_{i=1}^p H(\eta_i) \right] + A \sum_{i=1}^p \sum_{j \neq i}^p [\eta_i^2(\mathbf{r})\eta_j^2(\mathbf{r})]. \tag{9}$$

$H(\eta_i)$ is an interpolation function to connect $f_\alpha$ and $f_{\theta'}$, which is given by $H(\eta_i) = \eta_i^3(10 - 15\eta_i + 6\eta_i^2)$. It satisfies the constrains that $H(0) = 0$, $H(1) = 1$ and $dH(\eta_i)/d\eta_i = 0$ at $\eta_i = 0$ and 1. $A$ is a constant that is determined by domain wall energy between different θ' variants. A dimensionless value of 10 was chosen in the current study, so the domain-wall energy is more than twice higher than the interfacial energy between the α-Al/θ' phases to prevent the coalescence of different θ' variants.

*3.2 Anisotropic interfacial energies*

The coherent interface is (001)$_{\theta'}$/(001)$_\alpha$ for the θ' variant whose lattice is shown in Fig. 1, whereas the (100)$_{\theta'}$/(100)$_\alpha$ and (010)$_{\theta'}$/(010)$_\alpha$ interfaces are semi-coherent [3, 24, 29]. According to MS simulations, the energies of the coherent and semi-coherent interfaces are 156 mJ/m$^2$ and 694 mJ/m$^2$, respectively [24]. In this work, the anisotropic interfacial energy is incorporated in the third rank tensor $\beta_{ij}(p)$. For the variant shown in Fig. 1, and assuming that [100]$_\alpha$, [010]$_\alpha$ and [001]$_\alpha$ correspond to the $x$, $y$ and $z$ directions, $\beta_{ij}(p)$ is given by [20, 29]:

$$\beta_{ij}(p) = \begin{pmatrix} 40.85 & 0 & 0 \\ 0 & 40.85 & 0 \\ 0 & 0 & 1.50 \end{pmatrix}. \tag{10}$$

The numbers in Eq. (9) are non-dimensional and they were chosen together with the gradient coefficient of the concentration field, $\kappa_c = 15$, to ensure that the interfacial energies of the coherent and semi-coherent interfaces are consistent with the MS calculation results and to avoid the artificial fraction. The energy of the matrix/precipitate interface, $\gamma(s)$, depends on the orientation of the precipitate and can be expressed as:

$$\gamma(\mathbf{s}) = \sqrt{\left(\gamma_{(100)} \cos \lambda \cos \varphi\right)^2 + \left(\gamma_{(010)} \sin \lambda \cos \varphi\right)^2 + \left(\gamma_{(001)} \sin \varphi\right)^2}, \tag{11}$$

where $s$ is the normal vector to the interface and $\lambda$ and $\varphi$ stand for the angles between $s$ and the projection of $s$ on $(001)_\alpha$ and $[100]_\alpha$, respectively. $\gamma(s)$ is plotted in Fig. 3. It worth noting that semi-coherent interfaces are treated in this model as coherent interfaces with a larger interfacial energy.

*3.3 Elastic strain energy*

*3.3.1 Microelasticity theory*

The elastic strain energy associated with the shear transformation strain is formulated using the Khachatruyan and Shatalov's microelasticity theory [38]. A stress-free boundary condition is applied in this investigation [32, 39-40], and the elastic strain energy is given by:

$$E^{elas} = \frac{1}{2} \sum_{p,q=1} \oint \frac{d^3\mathbf{g}}{(2\pi)^3} B_{pq}\left(\frac{\mathbf{g}}{|\mathbf{g}|}\right) \{\tilde{\eta}_p\}_\mathbf{g} \{\tilde{\eta}_q\}_\mathbf{g}^*, \tag{12}$$

where the integral is taken in the reciprocal space and **g** is a vector in the reciprocal space. Note that **g** = 0 is excluded from the integration, which defines the principle value. $\{\tilde{\eta}_p\}_\mathbf{g}$ is the Fourier transform of $\eta(\mathbf{r})$. The asterisk indicates the complex conjugate. Considering a system with the stress-free boundary condition, $B_{pq}(\mathbf{g}/|\mathbf{g}|)$ can be expressed as [38]:

$$B_{pq}\left(\frac{\mathbf{g}}{|\mathbf{g}|}\right) = B_{pq}(\mathbf{n}) = C_{ijkl}\varepsilon_{ij}(p)\varepsilon_{kl}(q) - n_i\sigma_{ij}(p)\Omega_{jk}\sigma_{kl}(q)n_l \quad \mathbf{g} \neq 0, \tag{13}$$

where $\mathbf{n} = \mathbf{g}/|\mathbf{g}|$ is a unit vector in a reciprocal space, $\Omega_{ij}^{-1} = C_{ijkl}n_k n_l$ and $\sigma_{ij}(p) = C_{ijkl}\varepsilon_{kl}(p)$. $C_{ijkl}$ is the stiffness tensor of the matrix. The homogeneous modulus approximation is assumed in the simulations [37] and therefore the values of $C_{ijkl}$ for the α-Al matrix ($C_{11}$ = 104.4 GPa, $C_{12}$ = 59.6 GPa and $C_{44}$ = 28.1 GPa) calculated using DFT were used [23]. They are very close to the actual values of α-Al experimentally measured [40]. The scaling factor between the actual elastic constants and dimensionless elastic constants is 1, according to [31], and the dimensionless values of $C_{ijkl}$ used in equation (13) were $C_{11}$ = 104.4, $C_{12}$ = 59.6 and $C_{44}$ = 28.1. $\varepsilon_{ij}(p)$ stands for the stress free transformation strain tensor (SFTS) of the $p$th variant of the θ' precipitates, which is obtained from the deformation gradient matrix $T$ via:

$$\varepsilon_{ij}(p) = \frac{T_p^T T_p - I}{2}, \tag{14}$$

where $T^T$ is the transpose matrix of $T$ and $I$ is the identity matrix. $T$ is derived from the lattice transformation form α-Al to θ' (see Fig. 1). For instance, the form of $T$ for the variant shown in Fig. 1, according to the lattice correspondence proposed by Nie *et al.* [3, 28], is given by:

$$T_{①} = \begin{pmatrix} \frac{a_{\theta'}}{a_{\alpha}} & 0 & 0 \\ 0 & \frac{a_{\theta'}}{a_{\alpha}} & -\frac{1}{3} \\ 0 & 0 & \frac{2c_{\theta'}}{3a_{\alpha}} \end{pmatrix}. \tag{15}$$

where $a_{\theta'}$, $c_{\theta'}$ and $a_{\alpha}$ are the lattice parameters of θ′ and α-Al matrix. The transformation matrices of the other θ′ variants can be obtained via symmetry operations, as it will be shown in the results section.

*3.3.2 Calculation of the lattice parameters*

The lattice parameters θ′ and α-Al matrix were calculated via atomic-scaled DFT simulations, which were preformed using the Vienna Ab-initial Simulation Package (VASP) [42–43] in a non-spin-polarized condition. The projected-augmented-wave (PAW) method [44–45] and the Perdew-Burke-Ernzerhof (PBE) generalized gradient approximation (GGA) [46–47] were adopted. The cut-off energy of the plane wave basis set was 350 eV. The $k$-space integrations were performed using a Monkhorst Pack sampling scheme [48], with an equivalent (18 × 18 × 18) $k$-point mesh. Geometry optimization was performed until the total energy change was less than $10^{-3}$ eV/atom.

*3.4 Interaction energy*

The influence of external stresses on the precipitation of the $p$th variant of the θ′ precipitate is included via an extra elastic energy interaction term, $E^{int}$, given by [49–52]:

$$E_p^{int} = -\int \sigma_{ij}^{ex}(\mathbf{r}) \sum_{1}^{p} \varepsilon_{ij}(p)\eta(\mathbf{r})\, d\mathbf{r}, \tag{16}$$

where $\sigma_{ij}^{ex}$ is the external stress field and $\varepsilon_{ij}(p)$ stands for the SFTS of the $p$th variant of the θ′ precipitates. The interaction energy (density) between the external stress field and the SFTS of θ′ precipitates, which reveals the additional contribution of the external stress to the formation of θ′, is evaluated as the variational derivative of $E^{int}$ to $\eta_p$ [49–52]:

$$e_p^{int} = \frac{\delta E^{int}}{\delta \eta_p} = -\sigma_{ij}^{ex}(\mathbf{r})\varepsilon_{ij}. \tag{17}$$

The nucleation of θ′ variants with negative interaction energies will be favoured. It should be noticed that $e_p^{int}$ depends on the external stress field and the SFTS associated with each variant but it is independent of the size and shape of the precipitate.

This interaction energy was used in this work to assess the effect of pre-existing dislocations on the formation of the θ′ precipitates. In this case, $\sigma_{ij}^{ex}$ corresponds the stress field of a dislocation which is calculated based on the phase-field dislocation model developed by Wang et al. [53–55]. In this model, the eigenstrain field of a dislocation $\varepsilon_{ij}^{dis}(\mathbf{r})$ is expressed as $\varepsilon_{ij}^{dis\_0}\varphi(\mathbf{r})$, where φ is the order parameter of the dislocation and $\mathbf{r}$ is a vector in real space. $\varepsilon_{ij}^{dis\_0}$ is the SFTS of a dislocation, which is expressed as a dyadic product of the slip plane normal $\mathbf{n}$ and the Burgers vector $\mathbf{b}$ [50, 52, 56]:

$$\varepsilon_{ij}^{dis\_0} = \frac{\mathbf{b} \otimes \mathbf{n} + \mathbf{n} \otimes \mathbf{b}}{2d} = \frac{(b_i n_j + b_j n_i)}{2d}, \tag{18}$$

where $d$ is the interplanar spacing between adjacent slip planes.

*3.5 Numerical integration of the kinetics equations*

The Cahn-Hilliard, Eq. (1) and Allen-Cahn, Eq. (2), equations were solved by the semi-implicit Fourier-spectral method using periodic boundary conditions in a cubic domain. Following [56], the semi-implicit forms of time integration of the Cahn-Hilliard and Allan-Cahn equations in reciprocal space are given by (for the θ' variant shown in Fig. 1):

$$\tilde{c}^{t+1} = \frac{\tilde{c}^t - \Delta t \cdot M \cdot g^2 \left(\frac{\partial \tilde{f}^p}{\partial c}\right)_{\mathbf{g}}}{1 + \Delta t \cdot M \cdot g^4 \cdot \kappa_c}, \qquad (19)$$

$$\tilde{\eta}^{t+1} = \frac{\tilde{\eta}^t - \Delta t \cdot L \cdot \left(\frac{\partial \tilde{f}^p}{\partial \eta}\right)_{\mathbf{g}}}{1 + \Delta t \cdot L \cdot \left(g_x^2 \cdot \beta_{11} + g_y^2 \cdot \beta_{22} + g_z^2 \cdot \beta_{33}\right)}, \qquad (20)$$

where $\tilde{c}$ and $\tilde{\eta}^t$ are the Fourier transformed forms of $c$ and $\eta$, respectively. $\Delta t$ is the time step and $g_x$, $g_y$ and $g_z$ are the $x$, $y$ and $z$ coordinates of $\mathbf{g}$, $\tilde{f}^p$ is the total energy excluding the gradient energy. More details about the numerical integration technique can be found in [57].

## 4. Experimental techniques and results

*4.1 Experimental methods*

An Al - 4 wt. % Cu (1.7 at. %) alloy was prepared using high- purity metals by induction casting in an induction furnace (VSG 002 DS, PVA TePla). The pure metals were melted using an alumina crucible in an inert Ar atmosphere. The molten alloy was homogenized during 15 minutes and poured into a stainless steel mould with cylindrical cavities Samples were machined from the central region of the ingot and subjected to a solution heat treatment during 22 h at 540ºC. They were quenched in water and aged during 50 and 66 h at 180ºC. Specimens for observation in the transmission electron microscope (TEM) were gently ground to a thickness of approximately 100 µm, and jet electropolished with a solution of 30% of nitric acid and 70% of methanol (vol. %) at ≈ -30ºC. Foils with a thickness of about 100 nm were analysed in a FEI Talos TEM at 200 kV in a high-angle annular dark-field imaging scanning transmission electron microscope mode (HAADF-STEM) in the $\langle 100 \rangle_\alpha$ orientation. In this orientation, the habit planes of two variants of the θ' precipitates were parallel to the electron beam. The foil thickness in the beam direction is determined by measuring the spacing of Kossel-Möllenstedt fringes in a $(022)_\alpha$ reflection in a $\langle 100 \rangle_\alpha$ two-beam convergent beam electron diffraction pattern. The thickness and diameter of θ' precipitates were measured manually, using the TEM Image Analysis Offline software, in approximately 300 particles.

*4.2 HAADF-STEM observations*

HAADF-STEM images revealed the distribution of θ' precipitates because the contrast is related to the atomic number and θ' precipitates appear brighter than α-Al since the θ' phase has higher Cu content than the α-Al matrix (Fig. 4). Samples were oriented perpendicular to the $[001]_\alpha$ axis and the precipitates with habit planes $(100)_{\theta'}$ and $(010)_{\theta'}$ are clearly seen. Our HAADF-STEM images, as well as previous observations in the literature [2, 3, 25], show that most of θ' precipitates are randomly distributed within the α-Al matrix (Fig. 4a). Moreover, θ' precipitates with the same habit plane stacked into a parallel array were also found, which is consistent with previous observations [2, 27]. Smaller θ" precipitates were also present. They were distinguished from θ' precipitates using higher magnification imaging, because θ" precipitates are formed by 2 Cu layers encompassing 3 Al planes.

## 5 Results and discussion

*5.1 Deformation variants of θ'*

The lattice parameters of θ' and α-Al computed by DFT are $a_{θ'} = 0.4080$ nm, $c_{θ'} = 0.5701$ nm and $a_α = 0.4050$ mn, which are close to the experimental results in [3]: $a_{θ'} = 0.404$ nm, $c_{θ'} = 0.580$ nm and $a_α = 0.404$ nm. Based on the computed lattice parameters, the deformation gradient matrix $T$ of the variant shown in Fig. 1 (denominated variant ① from here on) is given by

$$T_① = \begin{pmatrix} 1.0077 & 0 & 0 \\ 0 & 1.0077 & -0.3333 \\ 0 & 0 & 0.9384 \end{pmatrix}. \tag{21}$$

The dilatational strain along $[100]_α$ and $[010]_α$ (0.0077) is much lower for this variant than the dilatational strain along $[001]_α$ (0.0480) and the shear strain (1/3). In addition, this dilatational strain will be 0 if the experimental lattice parameters of α-Al and θ' are used. Thus, it is reasonable to ignore this small strain (0.0077) in the analyses, and $T$ and $ε_{ij}$ of the variant ① can be expressed as:

$$T_① = \begin{pmatrix} 1 & 0 & 0 \\ 0 & 1 & -0.3333 \\ 0 & 0 & 0.9384 \end{pmatrix}, \quad ε_① = \begin{pmatrix} 0 & 0 & 0 \\ 0 & 0 & -0.1667 \\ 0 & -0.1667 & -0.0042 \end{pmatrix} \tag{22}$$

The possible habit plane(s) of this variant, corresponding to $T_①$, can be calculated as [58–59]:

$$Q_① U_① - I = \boldsymbol{a}_① \times \boldsymbol{n}_①, \tag{23}$$

where $Q$ is the rigid-body rotation matrix and $U = \sqrt{T'T}$. $\boldsymbol{a}$ stands for the lattice vector and $\boldsymbol{n}$ is the habit plane normal. Two possible habit planes emerge from solving Eq. (22) and their normal vectors are expressed as:

$$\boldsymbol{n}_{①-1} = \begin{bmatrix} 0 \\ 0 \\ 1 \end{bmatrix} \quad \text{and} \quad \boldsymbol{n}_{①-2} = \begin{bmatrix} 0 \\ 0.9998 \\ 0.0125 \end{bmatrix}. \tag{24}$$

The first habit plane is $(001)_{θ'}$ (see Fig. 1), which corresponds to a $(001)_{θ'}/(001)_α$ interface, while the second habit plane is irrational. The θ'/α-Al interface corresponding to this possible irrational habit plane leads to a large atomic mismatch, which will result in a much higher interfacial energy. Thus, this habit plane and its corresponding interface are not realistic from the physical viewpoint.

There are 12 possible transformations of α-Al → θ' because the habit plane of variant ①, i.e., $(001)_α$, has the 4-fold-symmetry and 3 crystallographic equivalent planes: $(001)_α$, $(100)_α$ and $(010)_α$. These 12 transformation modes relate to 12 α-Al → θ' lattice correspondences (LC) (see Table 1), and thus to 12 deformation variants (DVs, marked as ①, ②, ..., ⑫) [60–61]. In the phase-field simulation, each structural order parameter $η_p$ stands for the corresponding $p$th DV. The deformation gradient matrices, $T_p$, and the corresponding SFTS, $ε_p$ of the 12 DVs are listed in Table 1 and can be easily derived from $T_①$ through the appropriate symmetry operations.

The 12 DVs can be divided into 3 groups: ①–④, ⑤–⑧, and ⑨–⑫. In each group, the four DVs have the same type of α-Al/θ' orientation relationship (OR, see Table 1) and these four DVs correspond to one orientation variant (OV) [59]. This indicates that the θ' phase has three OVs, which is consistent with the analyses via point groups. In terms of point groups, the number of OV is determined by the order of the point group of the parent phase, i.e., the matrix, divided by the order of the intersection group between the parent and product phases at a given OR [62]. In the α-Al → θ' transformation, the point group of the product phase (θ') is $4/mmm$, which is a subgroup of that of the parent phase (α-Al, $m\bar{3}m$). The order of the $4/mmm$, which is the intersection group, is 16 whereas that of the $m\bar{3}m$ is 48. Thus, the number of OV of θ' is 48/16 = 3.

*5.2 Shape of θ' precipitates*

Phase-field simulations of the formation of θ' in α-Al were carried out in a cubic cell of dimensions $256l_0 \times 256l_0 \times 256l_0$ for an Al - 1.74 at. % Cu alloy aged at 200 °C. According to the interfacial energies obtained in Section 3.2, the grid spacing of the phase-field simulation is determined to be $l_0$ = 1.2 nm, and thus the size of the cell is 307 nm × 307 nm × 307 nm. In order to determine the precipitate size, it is assumed that a grid point belongs to the *p*th DV of the precipitate if $\eta_p > 0.5$. Otherwise it is assumed to belong to the matrix.

The equilibrium shape of a θ' precipitate (DV ⑧) obtained from the phase-field simulation depends on the energy contributions taken into account in the model, which can include the chemical free energy (isotropic), the interfacial energy and the elastic energy associated to a pure dilatational strain and to a combination of dilatational and shear strains for the transformation. They are plotted and compared in Fig. 5. Of course, the precipitate shape is spherical if the only contribution is the isotropic chemical free energy and no coherency elastic energy is considered (Fig. 5a). If the anisotropic contribution of the interface energy is included in the simulations, the precipitate has an ellipsoidal shape with an average aspect ratio approximately to 4:1 and a $(010)_{θ'}$ habit plane (Fig. 5b), which agrees with the Wulff plot [63]. The aspect ratio of the ellipsoidal precipitate increased to approximately 9:1 when the lattice dilatation (~0.048 compression strain along $[010]_{θ'}$) was included in the simulation (Fig. 5c). Notice that the aspect ratio of a precipitate depends on the particle size once the anisotropy of the elastic strain energy is considered because the elastic energy contribution becomes dominant with respect to the interfacial energy when the particle size increases. Finally, if the shear transformation strain (1/3) is included in the analysis, the precipitate reaches to a plate-like shape at equilibrium, with an aspect ratio around 23:1 and an average diameter of ~ 180 nm. These results show the interplay of the different energy contributions to select the growth habits and the shape of the precipitates. For instance, both interfacial and elastic strain energies prefer this precipitate variant to grow in $(010)_{θ'}$ because the $(010)_{θ'}/(010)_α$ interface has the lowest interfacial energy, the lattice dilation strain supresses the growth of θ' precipitates perpendicular to $(010)_{θ'}$ and the shear strain facilitates the growth along $(010)_{θ'}$ as well because the $(010)_{θ'}$ is an invariant plane in the α-Al → θ' transformation.

The simulated shapes of all 12 DVs are depicted in Fig. 6. This figure reveals that the habit planes of DVs ① – ④, ⑤ – ⑧ and ⑨ – ⑫ are $(001)_α$, $(010)_α$ and $(100)_α$, respectively, which is consistent with the previous crystallography analyses (Table 1). Notice that the broad faces of the precipitates are not equiaxed. For example, the length along $[010]_α$ for DV ① (Fig. 6a) is longer than that along $[100]_α$. This difference can be explained using an analogy between a θ' precipitate and a dislocation loop because the shear strain (~0.33) associate with the θ' precipitate is far larger than the lattice dilatation strain (~0.048). Thus, the eigenstrain tensor of both a dislocation and a θ' precipitate θ' are shear-dominated, and the stress field and elastic strain energy of a dislocation loop is similar to that of a precipitate plate [54–55]. Dislocation loops are not circular because the elastic strain energy (density) of a screw segments is lower than that of the edge segments with the same Burgers vector [64]. The shear direction of DV ① is along $[010]_α$, and thus the upper and lower fringes of the precipitate in Fig. 5a are analogous to the screw dislocation segments while the left and right fringes are equivalent to the edge segments.

The equilibrium shape of the θ' precipitate (as given by the average diameter and the thickness) was computed for precipitates of different size by changing the initial composition of the supersaturated solid solution. Only one precipitate was contained in the simulation cell in this case for convenience [20]. The simulation results are compared in Fig. 7 with the experimental data measured by HAADF-STEM in the Al – 4wt.% Cu alloy aged at 180ºC as well as with other data available in the literature for this system [65, 66]. The diameter and thickness of θ' precipitates determined from the phase field simulations were in very good agreement with the experimental data when the diameter of the θ' precipitates were in the range 85 nm and 200 nm and it should be noted that there are no fitting parameters in the multiscale model. All model inputs were obtained from DFT and MS simulations and the CALPHAD databases. The phase field simulations could not be used to determine the precipitate shape for very small precipitates (40 nm in diameter and 2 nm in thickness) because the limited resolution of the numerical grid used to solve the equations.

*5.3 Spatial distribution of θ' precipitates*

The spatial distribution of the θ' precipitates as a result of homogeneous and heterogeneous nucleation is analysed using the multi-scaled model constructed in this work. In the latter, the formation of θ' precipitates on pre-existing edge, screw and mixed dislocations with an $a/2\,[\bar{1}10]_\alpha$ Burgers vector is investigated.

*5.3.1 Random distribution of homogeneously nucleated θ' precipitates*

Random precipitate distribution due to homogeneous nucleation can be simulated in the phase-field framework using the Langevin noise terms in Eqs. (1) and (2) [49, 50]. The simulation results are depicted in Fig. 8, which shows the random distribution of plate-like shaped θ' precipitates. Six of them have $(100)_\alpha$, $(010)_\alpha$ and $(001)_\alpha$ habit planes and the average diameter and thickness of these precipitates were ≈ 200 nm and ≈ 9 nm, respectively. These are consistent with experimental observations in Figs. 4 and 7. All three OVs of θ' precipitates are produced by the simulations, and they are distinguished by their habit planes. 5 out of 12 DVs (①, ④, ⑦, ⑨ and ⑫) can be found in this simulation, as indicated in the figure.

*5.3.2 Heterogeneous nucleation of θ' precipitates on dislocations*

*5.3.2.1 Precipitation on an $[1\bar{1}0](110)$ edge dislocation*

The heterogeneous nucleation and growth of θ' precipitates on an edge dislocation was first analysed. The location of the pre-existing edge dislocation in the cubic domain is shown in Fig. 9a. The Burgers vector **b** of the dislocation is $a/2\,[\bar{1}10]_\alpha$, and the dislocation line, given by the vector $\xi$, is parallel to $[\bar{1}\bar{1}2]_\alpha$. The extra half plane of atoms of the edge dislocation is located above the slip plane, which is coloured by pink. All three OVs of θ' precipitates are produced by the simulations, and they are distinguished by their habit planes. 4 out of 12 DVs (⑥, ⑧, ⑩ and ⑫) can be found in this simulation, as indicated in the figure.

At time step $t = 1500$ in the simulation, the distribution of θ' precipitates nucleated around the edge dislocation is shown in Fig. 9b. All precipitates are plate-like shaped and they are filled with different colours to distinguish different DVs. These precipitates present $(010)_\alpha$ or $(100)_\alpha$ habit planes, which implies that precipitates with 2 different OVs (❷ and ❸, Table 1) grew on the given edge dislocation. Four arrays of θ' precipitate are formed along the dislocation line and the precipitates in each array belong to the same DV. The first array is located above the $(111)_\alpha$ slip plane, and is formed by four stacked θ' precipitates (DV ⑥, navy blue) that have a $(010)_\alpha$ habit plane. θ' precipitates with the same habit plane that belong to the DV ⑧ form another array along the dislocation line below the slip plane. Two precipitates that belong to the DV ⑫, and another two from the DV ⑩ are formed to the left and to the right of the dislocation line, respectively, with a $(100)_\alpha$ habit plane. The volume fractions of DVs ⑥ and ⑫ are 2.22% and 1.90%, respectively, which are higher than those of DVs ⑧ and ⑩ (0.62% and 0.50%).

The intersection of the precipitates nucleated on the dislocations with the $(100)_\alpha$ and a $(010)_\alpha$ planes are depicted in Figs. 9c and d, respectively, to make easier the comparison with experimental observations since the TEM images provide a 2D projection. The α-Al matrix is shown in light blue in these two figures, while θ' precipitates are shown in red. The cross-section of three θ' precipitates (DV ⑥) are seen in Fig. 9c, while two θ' precipitates (DV ⑫) are observed in Fig. 9d. The intersected precipitates in Figs. 9c and 9d present a similar spatial distribution: they are nearly parallel and are stacked in arrays along $[001]_\alpha$, which is consistent with the results found in HAADF-STEM images (Fig. 4b).

The selection of the DVs that nucleate and grow on the edge dislocation is dictated by the interaction energy, $E_p^{int}(\mathbf{r})$, between the stress field of the edge dislocation and the SFTS of the 12 DVs of θ' precipitates, and the total elastic strain energy. The contribution of the elastic strain energy to nucleation is negligible due to the small precipitate size and the interaction energy plays the dominant role. The minimum values of the interaction energies, $\min(e_p^{int})$, for all the DVs are plotted in Fig. 10. They are all negative, i.e. the stress field of the pre-existing edge dislocation facilitates the nucleation of precipitates on the dislocation. Among them, two DVs (⑥ and ⑫) stand out as the most energetically favourable to be nucleated on the dislocation, followed by DVs ⑧ and ⑩.

The interaction energy field around the dislocation in a $(001)_\alpha$ intersection are plotted in Figs. 11a to 11d for DVs ⑥, ⑧, ⑩, and ⑫, respectively, to understand the locations around the dislocation in which the precipitates from each DV were nucleated (Fig. 9b). In each figure, the position of the edge dislocation is represented by "⊥" symbol and the corresponding $(111)_\alpha$ slip plane is presented by the purple line. The shape of the to-be-nucleated θ' variants are schematically shown by a dashed rectangle in each figure. The most negative interaction energies are shown in blue and stand for the most favourable locations for the nucleation of each DV. They are found above the slip plane for DVs ⑥ and ⑫ (Figs. 11a and 11b), and below the slip plane for DVs ⑧ and ⑩ (Figs. 11c and 11d). These results are consistent with the simulation results shown in Fig. 9b. It should be noted that only one variant is allowed to nucleate in each location (see the last term in Eq. (8)). Thus, once these four DVs nucleate and grow along the dislocation line, the precipitate arrays form and the formation of the other DVs are penalized.

Once one precipitate variant is nucleated, the growth is controlled by the contributions of the interaction ($E_p^{int}$) and elastic ($E^{elas}$) energies. They can be ascertained by the comparison of the mean stress field, $(\sigma_{xx} + \sigma_{yy} + \sigma_{zz})/3$, induced by the SFTS of each DV with that induced by the edge dislocation. The contour plots in the $(001)_\alpha$ plane of the mean stress fields associated the DVs ⑥ and ⑧ are plotted in Figs. 12a and 12b, respectively, while the mean stress field of the edge dislocation is depicted in Fig. 12c. According to Fig. 11a, DV ⑥ nucleates above the slip plane. Comparison of the mean stress fields in Figs. 12a and 12c show that the tensile and compressive stress fields at the lower left and right sites of the precipitate (Fig. 12a) can be compensated by the compressive and tensile stress fields on the left and on the right of the dislocation core. Thus, DV ⑥ is energetically favoured to grow along the vertical habit plane (Fig. 11a). The situation of DV ⑧ is similar, and this variant will form below the slip plane and grow along the vertical habit plane. However, the volume fraction of DV ⑥ is significantly larger than that of DV ⑧ for two reasons. Firstly, interaction energy of DV ⑥ is more negative than that of DV ⑧. Secondly, the SFTS of both variants also includes a compressive strain of ~5% - 6% perpendicular to the broad faces of the precipitate generated by the lattice dilation, which results in a tensile stress field in the matrix around the broad face of precipitate (Figs. 12d and 12e). This tensile stress field can be compensated by the compressive stress field induced by the edge dislocation above the glide plane if the precipitate grows in this region, as it is the case of DV ⑥. If the precipitate nucleates and grows below the slip plane of the edge dislocation (as it is the case of DV ⑧ shown in Fig. 9b), the growth of the precipitate is hindered by the tensile stress generated by the edge dislocation below the slip plane.

*5.3.2.2 Precipitation on an [1$\bar{1}$0](110) screw dislocation*

The position of the pre-existing screw dislocation with $\boldsymbol{b} = a/2\,[\bar{1}\bar{1}0]_\alpha$ in the slip plane $(111)_\alpha$ is shown in Fig. 13a. The simulation results of the precipitate distribution around the screw dislocation after ageing at 200 ºC at $t = 1500$ are shown in Fig. 13b. In this figure, the slip plane is indicated in light pink and different DVs of the θ' precipitates are shown different colours. It is readily seen that all θ' precipitates have a plate-like shape and are distributed along the dislocation line. Three θ' precipitates (DV ⑨, sky blue) are located above the slip plane in the centre of the simulation box, while several stacked precipitates (DV ⑤, purple) appear behind the slip plane. Finally, one θ' plate that belongs to DV ⑦ (navy blue) was also found.

The distribution of θ' precipitates can be more clearly observed when viewed from $[001]_\alpha$ (Fig. 13c). The screw dislocation in this view appears as the purple line, and the Burgers vector of the dislocation is marked by an arrow. The $(111)_\alpha$ slip plane is coloured in pink. The θ' precipitate plates have either a $(010)_\alpha$ or a $(010)_\alpha$ habit plane, i.e. belong to the OVs ❷ and ❸. Within these orientations variants, these precipitates also belong to 4 DVs: ⑤, ⑦, ⑨ and ⑪. The three precipitates of DV ⑨ with $(100)_\alpha$ habit plane stack in an array on the right side of the screw dislocation. Some DV ⑦ precipitates with a $(010)_\alpha$ habit plane form between the neighbouring DV ⑨ precipitates. In addition, two other DV ⑦ precipitates are distributed on the right of the DV ⑨ array. The majority of the θ' precipitates stacked on the left site of the dislocation belong to the DV ⑤ with a $(010)_\alpha$ habit plane. A small θ' precipitate, that belongs to DV ⑪, is found on the upper right site of the array, which was covered by the other precipitates in Fig. 13b. The volume fraction of this variant (~0.48%) is much smaller than that of the other three variants, whose volume fractions are approximately 1.30%, 0.97% and 1.50% for DVs ⑤, ⑦ and ⑨.

The selection of the DVs that nucleate on the screw dislocation is again dictated by the minimum interaction energy, $\min(e_\mathrm{p}^\mathrm{int})$, which is plotted in Fig. 14 for the 12 DVs. In this case, the $\min(e_\mathrm{p}^\mathrm{int})$ values of DVs ① - ④ are closed to those of DVs ⑤, ⑦, ⑨ and ⑪, but DVs ① - ④ have not been observed in Fig. 13. This can be understood from the SFTSs of these DVs that are shown in Table 1. If the SFTS of DVs ① and ⑦, ② and ⑪, ③ and ⑤, and ④ and ⑨ are compared, the only difference is the position of the lattice dilation strain. The lattice dilation strain (-0.0042) is much smaller than the shear strain, leading to similar values of the $\min(e_\mathrm{p}^\mathrm{int})$ values between pairs of DVs (for instance ④ and ⑨) which have the same shear strain. But each pair of these DVs have different orientation relationships with the α-Al matrix, which results in a different habit plane (the habit planes of DVs ④ and ⑨ are $(100)_\alpha$ and $(001)_\alpha$, respectively). Thus, the stress field around DV ④ is different from that of DV ⑨, which result in the elastic interaction between the screw dislocation and DV ④ is different from that and DV ⑨.

Unlike the edge dislocation, the mean stress around a screw dislocation is 0. In this case, the contour plot of the deviatoric (Von Mises) stress field induced by the screw dislocation in the $(\bar{1}10)_\alpha$ plane (perpendicular to the dislocation line) is plotted in Fig. 15a. The same fields can be found in Figs. 15b and 15c, which is superimposed by the Von Mises stress fields of DVs ⑨ and ④, respectively. The maximum stress around the dislocation core in Fig. 15a is ≈4.8 GPa and decreases to ≈ 4.0 GPa in Fig. 15b and to ≈4.6 GPa in Fig. 15c. Thus, DV ⑨ is more effective in reducing the total elastic strain energy of the simulation system and its growth is favoured with respect to DV ④. Similar mechanisms explain why DVs ⑤, ⑦ and ⑪ develop instead of DVs ③, ① and ②.

*5.3.2.3 Precipitation on [1$\bar{1}$0]$_\alpha$(110)$_\alpha$ mixed dislocations*

The formation of θ' precipitates on two different mixed dislocations was simulated. Both dislocations are shown in Figs. 16a and b. The angles between the Burgers vector ***b*** and the line vector ***ξ*** of the two dislocations are 60° (Fig. 17a) and 30° (Fig. 16b), respectively. The distribution of the θ' precipitates formed on the mixed dislocation in Fig. 17a is shown in Fig. 16c. All precipitates have a (010)$_\alpha$ habit plane. 6 precipitates belonging to the DV ⑥ are formed and stacked into an array above the dislocation line. In addition, 2 precipitates of DV ⑧ form below the dislocation line. In the case of the θ' precipitates on the dislocation represented in Fig. 16b, they also have a (010)$_\alpha$ habit plane. These precipitates belong to 3DVs, i.e., ⑤, ⑥ and ⑧. DVs ⑤ and ⑧ are alternatively distributed above the slip plane while DV ⑥ forms below the slip plane. The dominant DVs of θ' precipitates formed on the mixed dislocations can be obtained following the analysis presented above for pure edge and scree dislocations and will not be discussed for the sake of brevity.

*5.4 Limitations of the model*

Nucleation of the precipitates within the phase-field framework is artificially triggered by introducing Langevin noise in eq. (1) and (2). In the absence of dislocations, this leads to the homogeneous nucleation of precipitates while heterogeneous nucleation is induced by the interaction energy between dislocations and precipitates. However, other factors that influence the nucleation of θ' precipitates (such as grain boundaries) are not considered in the model. In addition, the solute segregation at dislocations is ignored. Finally, the homogeneous modulus approximation is adopted in the phase field simulations although it is known that the elastic constants of the θ' phase are different from those of α-Al matrix. This approximation can be overcome in the future by taking into account the differences in the elastic constants between the matrix and the precipitate to calculate elastic strain energy [67].

# 6 Conclusions

A multiscale modelling strategy, based on the meso-scaled phase field approach, has been developed to simulate the morphology and spatial distribution of θ' precipitates in Al-Cu alloys. The model parameters that dictate the different energy contributions (chemical free energy, interfacial energy, lattice parameters) were obtained from the computational thermodynamics databases or from *ab initio* and molecular dynamics simulations, while the deformations associated with the transformation strain were obtained from geometrical considerations. Thus, with the exception of numerical parameters to ensure convergence, the multiscale model was free of empirical or adjustable parameters.

Homogeneous nucleation and growth of θ' precipitates at 200 ºC was simulated in a cubic domain. The model was able to reproduce the random distribution of three different orientation variants of the θ' precipitates with orientation relationship (001)$_{θ'}$//(001)$_\alpha$ and [100]$_{θ'}$//[100]$_\alpha$. The θ' precipitates have a plate-like shape with the broad surface parallel to the habit plane. The aspect ratio of the precipitates was controlled by the anisotropic interface energy and the elastic energy associated to the shear transformation strain. The aspect ratio was ≈4:1 if only the former was considered in the simulation, while the addition of the latter led to an aspect ratio of ~25:1, which is in consistent with the experiments.

The multiscale phase-field model was also used to simulate the heterogeneous precipitation at 200ºC on dislocations. If the precipitates form on an edge dislocation, 4 deformation variants appear and each array only contains one deformation variant of the precipitate with the same habit plane. Also 4 deformation variants appear if the θ' precipitates develop on a screw dislocation. The precipitates will present two habit planes in this case and precipitates with the habit plane normal perpendicular to the dislocation line will not form. The precipitates also stack into arrays on mixed dislocations.

The nucleation of θ' precipitate on the pre-existing dislocation is well predicted by interaction energy calculation results. The deformation variants that favoured by both the interaction energy and stress field are most favourable to nucleate along the dislocations. The stress/strain field analyses is used to determine whether a θ' precipitate is energetically favourable to form at a certain position near the pre-existing dislocation.

**Acknowledgements**

This investigation was supported by the European Research Council (ERC) under the European Union's Horizon 2020 research and innovation programme (Advanced Grant VIRMETAL, grant agreement No. 669141). BB acknowledges the support from the Spanish Ministry of Education through the fellowship FPU15/00403. Useful discussions with Prof. J.-F. Nie, Dr. Y. Gao, Prof. Y. and Dr. F. Lin are also gratefully acknowledged as well as the computer resources and technical assistance provided by the Centro de Supercomputación y Visualización de Madrid (CeSViMa).

**References**

[1] I.J. Polmear, Light Alloys: Metallurgy of the light metals, 3rd ed, Arnold, London, 1995.
[2] A. Kelly, R.B. Nicholson, Precipitation hardening, Prog. Mater. Sci. 10 (1963) 151–391.
[3] J.F. Nie, In D.E. Laughlin, K. Hono, editors. Physical Metallurgy, 5th ed, Elsevier, 2014.
[4] J.F. Nie, B.C. Muddle, Comments on the "dislocation interaction with semi–coherent precipitates (Ω phase) in deformed Al–Cu–Mg–Ag alloy", Scripta Mater. 42 (2000) 409–13.
[5] J.F. Nie, Effects of precipitate shape and orientation on dispersion strengthening in magnesium alloys, Scripta Mater. 48 (2003) 1009–15.
[6] J.F. Nie, B.C. Muddle, Strengthening of an Al–Cu–Sn alloy by deformation–resistant precipitate plates, Acta Mater. 56 (2008) 3490–501.
[7] H. Liu, Y. Gao, L. Qi, Y. Wang, J.F. Nie, Phase–field simulation of Orowan strengthening by coherent precipitate plates in an aluminium alloy, Metall. Mater. Trans. A 46 (2015) 3287–302.
[8] A.J. Kulkarni, K. Krishnamurthy, S.P. Deshmukh, R.S. Mishra, Effect of particle size distribution on strength of precipitation–hardened alloys, J. Mater. Res. 19 (2004) 2765–73.
[9] U.F. Kocks, A statistical theory of flow stress and work–hardening, Phil. Mag. 13 (1962) 541–66.
[10] A.J.E. Forman, M.J. Makin, Dislocation movement through random arrow of obstacle, Phil. Mag. 14 (1966) 911–24.
[11] M.F. Ashby, In A. Kelly, R.B. Nicholson, editors. Strengthening mechanism in crystals. Elsevier, Amsterdam, 1971.
[12] A. Guinier, Structure of age-hardened aluminium-copper alloys, Nature 142 (1938) 569–70.
[13] E. Hornbogen, E.A. Starke Jr., Theory assisted design of high strength low alloy aluminium, Acta Metall. Mater. 41 (1993) 1–16.
[14] R. Becker, On the formation of nuclei during precipitation, Proc. Phys. Soc. 52 (1940) 71–6.
[15] J.W. Gibbs, The scientific papers of J. Willard Gibbs, vol. 1, Dover, New York, 1961.
[16] J.W. Cahn, J.E. Hilliard, Free energy of a nonuniform system I. Interface free energy, J. Chem. Phys. 28 (1958) 258–67.
[17] L.Q. Chen, Phase-field models for microstructure evolution, Annu. Rev. Mater. Res. 32 (2002) 113–40.
[18] N. Moelans, B. Blanpain, P. Wollants, An introduction to phase-field modelling of microstructure evolution, CALPHAD, 32 (2008) 268–94.

[19] Y. Wang, J. Li, Phase field modelling of defects and deformation, Acta Mater. 58 (2010) 1212–35.
[20] H. Liu, Y. Gao, J.Z. Liu, Y.M. Zhu, Y. Wang, J.F. Nie, A simulation study of the shape of β' precipitates in Mg-Y and Mg-Gd alloys, Acta Mater. 61 (2013) 453–66.
[21] D.Y. Li, L.Q. Chen, Computer simulation of stress–orientated nucleation and growth of θ' precipitates in Al-Cu alloys, Acta Mater. 46 (1997) 2573–85.
[22] D.A. Porter, K.E. Easterling, Phase transformation in metals and alloys, 2nd ed, Chapman & Hall, London.
[23] C. Wolverton, First-principles prediction of equilibrium precipitate shapes in Al-Cu alloys, Phil. Mag. Lett. 79 (1999) 683–90.
[24] S.Y. Hu, M.I. Baskes, M. Stan, L.Q. Chen, Atomistic calculations of interfacial energies, nucleus shape and size of θ' precipitates in Al-Cu alloys, Acta Mater. 54 (2006) 4699–707.
[25] L. Bourgeois, J.F. Nie, B.C. Muddle, Assisted nucleation of θ' phase in Al-Cu-Sn: the modified crystallography of tin precipitates, Phil. Mag. 85 (2005) 3487–509.
[26] L. Bourgeois, C. Dwyer, M. Weyland, J.F. Nie, B.C. Muddle, The magic thicknesses of θ' precipitates in Sn-microalloyed Al-Cu, Acta Mater. 60 (2012) 633-44.
[27] U. Dahmen, K.H. Westmacott, Ledge structure and mechanism of θ' precipitate growth in Al-Cu, Phys. State. Sol. 80 (1983) 249–62.
[28] J.F. Nie, B.C. Muddle, The lattice correspondence and diffusional-displacive phase transformations, Mater. Forum 23 (1999) 23–40.
[29] V. Vaithyanathan, C. Wolverton, L.Q. Chen, Multiscale modelling of θ' precipitation in Al-Cu binary alloys, Acta Mater. 52 (2004) 2973–87.
[30] S.M. Allen, J.W. Cahn, A microscopic theory for antiphase boundary motion and its application to antiphase domain coarsening, Acta Metall. 27 (1979) 1085–95.
[31] C. Shen, PhD Thesis, the Ohio State University, 2004.
[32] Y. Wang, A.G. Khachatruyan, Three-dimensional field model and computer modelling of martensitic transformations, Acta Mater. 45 (1997) 759–73.
[33] C. Shen, J.P. Simmons, Y. Wang, Effect of elastic interaction on nucleation: II. Implementation of strain energy of nucleus formation in the phase field method, Acta Mater. 55 (2007) 1457–66.
[34] L.D. Landau, E.M. Lifshitz, Statistical Physics, Pergamon Press, Oxford, 1980.
[35] L.G. Zhang, L.B. Liu, G.X. Huang, H.Y. Qi, B.R. Jia, Z.P. Jin, Thermodynamic assessment of the Al-Cu-Er system, CALPHAD, 32 (2008) 527–34.
[36] O. Redlich, A.T. Kister, Algebraic representation of thermodynamic properties and the classification of solutions, Ind. Eng. Chem. 40 (1948) 345–8.
[37] H. Okamoto, P.R. Subramanian, L.P. Kacprzak, editors. Binary alloy phase diagrams. ASM International: Materials Park, Ohio; 1990.
[38] A.G. Khachatruyan, Theory of structural transformations in solids, 2nd ed, Dover, New York, 2008.
[39] N. Zhou, C. Shen, M.J. Mills, J. Li, Y. Wang. Modelling displacive-disffusional coupled dislocation shearing of γ' precipitates in Ni-base superalloys, Acta Mater. 2011, 59:3484–97.
[40] Y. Gao, N. Zhou, D. Wang, Y. Wang. Pattern formation during cubic to orthorhombic martensitic transformations in shape memory alloys, Acta Mater 2014; 68:93–105.
[41] J. Vallin, M. Mongy, K. Salama, O. Beckman. Elastic constants of aluminium, J. Appl. Phys. 1964; 35:1825–6.
[42] G. Kresse, J. Furthmuller, Efficient iterative schemes for ab inito total-energy calculations using a plane-wave basis set, Phys. Rev. B 54 (1996) 11169–11186.
[43] G. Kresse, D. Joubert, From ultrasoft pseudopentials to the projector augmented-wave method, Phys. Rev. B 85 (2012) 144301.
[44] P. Vostry, B. Smola, I. Stulikova, F. Buch, B.L. Mordike, Microstructure evolution in isochronally heat treated Mg–Gd alloys, Phys. Stat. Sol. A 175 (1999) 491–500.

[45] B. Smola, I. Stulikova, F. Buch, B.L. Mordike, Structural aspects of high performance Mg alloys design, Mater. Sci. Eng. A 324 (2002) 113–117.

[46] J.P. Perdew, Y. Wang, Accurate and simple analytic representation of the electron-gas correlation energy, Phys. Rev. B 45 (1992) 13244–13249.

[47] J.P. Perdew, K. Burke, M. Emzerhof, Generalized gradient approximation made simple, Phys. Rev. Lett. 77 (1996) 3865–3868.

[48] J.D. Pack, H.J. Monkhorst, "Special points for Brillouin-zone intergrations"- a reply, Phys. Rev. B 16 (1977) 1748–1749.

[49] C. Shen, J.P. Simmons, Y. Wang, Effect of elastic interaction on nucleation: I. Calculation of the strain energy of nucleus formation in an elastic anisotropic crystal of arbitrary microstructure, Acta Mater. 54 (2006) 5617–30.

[50] H. Liu, Y. Gao, Y.M. Zhu, Y. Wang, J.F. Nie, A simulation of $\beta_1$ precipitation on dislocations in an Mg-rare earth alloys, Acta Mater. 77 (2014) 133–50.

[51] Y. Gao, H. Liu, R. Shi, N. Zhou, Z. Xu, Y. M. Zhu, J.F. Nie, Y. Wang, Simulation study of precipitation in an Mg-Y-Nd alloy, Acta Mater. 60 (2012) 4819–32.

[52] D. Qiu, R. Shi, D. Zhang, W. Lu, Y. Wang. Variant selection by dislocations during α precipitation in α/β titanium alloys, Acta Mater. 88 (2015) 218–31.

[53] Y.U. Wang, Y.M. Jin, A. M. Cuitino, A.G. Khachatruyan, Nanoscale phase field microelasticity theory of dislocations: model and 3D simulations, Acta Mater. 49 (2001) 1847–57.

[54] Y.U. Wang, Y.M. Jin, A. M. Cuitino, A. G. Khachatruyan, Application of phase field microelasticity theory of phase transformations to dislocation dynamics: Model and three-dimensional simulations in a single crystal, Phil. Mag. Lett. 81 (2001) 385–93.

[55] Y.U. Wang, Y.M. Jin, A. M. Cuitino, A. G. Khachatruyan, Phase field microelasticity theory and modelling of multiple dislocation dynamics, Appl. Phys. Lett. 78 (2003) 2324–6.

[56] C. Shen, Y. Wang, Phase field model of dislocation networks, Acta Mater. 51 (2003) 2595–610.

[57] L.Q. Chen, J. Shen, Applications of semi-implicit Fourier-spectral method to phase field equations, Comp. Phys. Comm. 108 (1998) 147–58.

[58] K. Bhattacharya, Microstructure of martensite. Oxford University Press, New York, 2004.

[59] C.M. Wayman, Introduction to the crystallography of martensitic transformation, Collier-Macmillan, New York, 1964.

[60] Y. Gao, R. Shi, J.F. Nie, S.A. Dregia, Y. Wang, Group theory description of transformation pathway degeneracy in structural phase transformations, Acta Mater. 109 (2016) 353–63.

[61] Y. Gao, N. Zhou, D. Wang, Y. Wang, Pattern formation during cubic to orthorhombic martensitic transformations in shape memory alloys, Acta Mater. 68 (2014) 93–105.

[62] J.W. Cahn, G.M. Kalonji, Symmetry in solid-solid transformation morphologies. In: Proceedings of an international conference on solid-solid phase transformations; 1981.

[63] W.K. Burton, N. Cabrera, F.C. Frank, The growth of crystals and the equilibrium structure of their surfaces, Phil. Trans. Roy. Soc. Lond. Ser. A 866 (1951) 299–358.

[64] D. Hull, D.J. Bacon, Introduction to dislocations, 5th ed, Butterworth-Heinemann, Oxford.

[65] A.W. Zhu, J. Chen, E. A. Starke Jr, Precipitation strengthening of stress-aged Al-xCu alloys, Acta Mater. 48 (2000) 2239–2246.

[66] A. Biswas, D. J. Siegel, C. Wolverton, D. N. Seidman, Precipitates in Al-Cu alloys revisited: Atom-probe tomographic experiments and first-principles calculations of compositional evolution and interfacial segregation. Acta Mater. 59 (2011) 6187–6204.

[67] Y. U. Wang, Y. M. Jin, A.G. Khachaturyan, Phase field microelasticity theory and modelling of elastically and structurally inhomogeneous solid, J. Appl. Phys. 92 (2002) 1351–60.

**FIGURES**

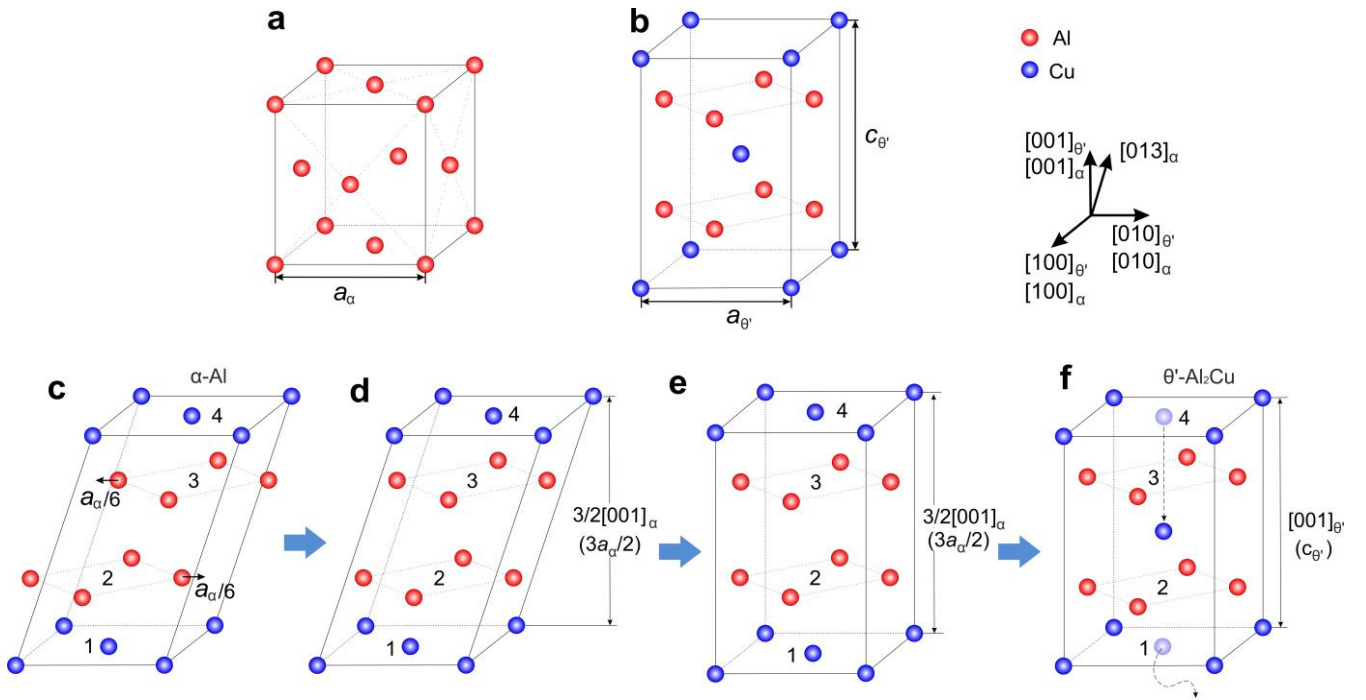

Fig. 1 Unit cells of (a) α-Al and (b) θ'. (c-f) shown the transformation pathway from the lattice of α-Al to that of θ'. Red and blue spheres represent Al and Cu atoms, respectively.

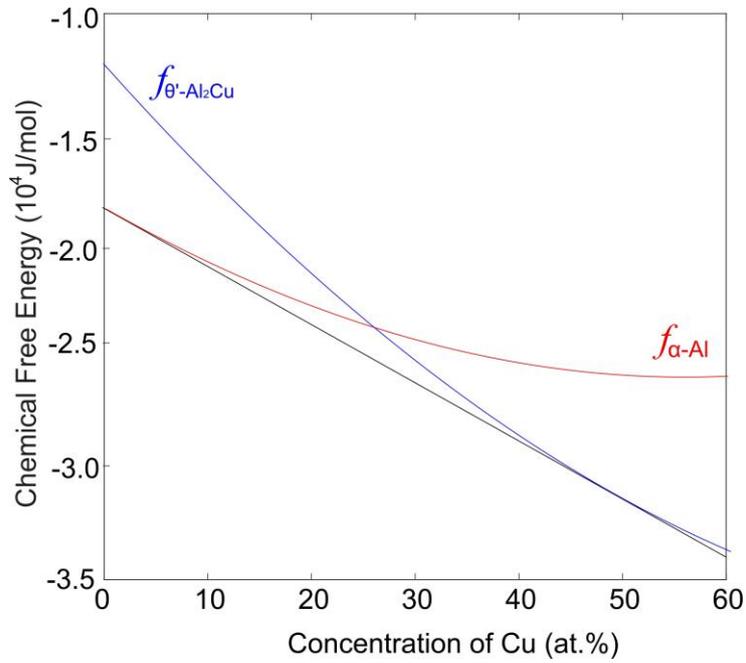

Fig. 2 Variation of chemical free energy of θ' (blue curve) and of α-Al (red curve) as a function of Cu concentration (at.%). The black line is the common tangent line between the chemical free energy curves of θ' and α-Al.

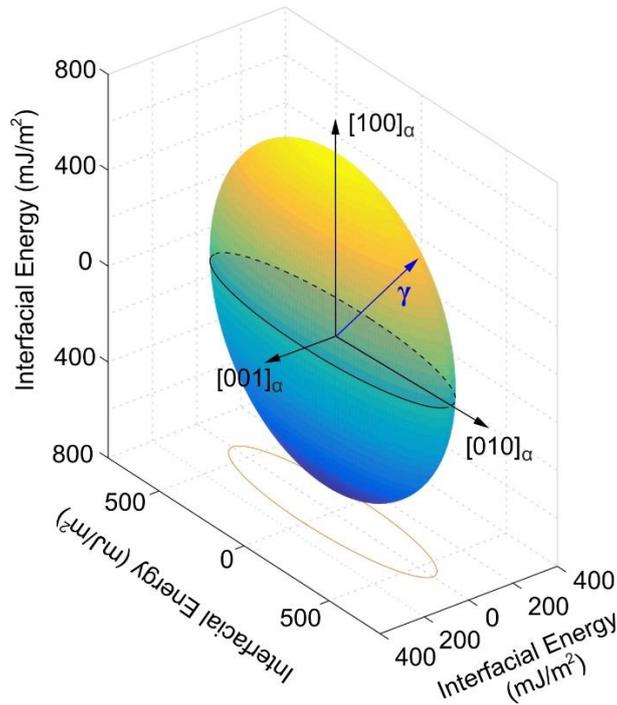

Fig. 3 Variation of the θ'/α-Al interfacial energy, γ(r), as a function of the normal vector to the interface. The interface normal vector is represented by the blue arrow, which starts from the center and ends on the surface of the ellipsoid. The length the vector represents the magnitude of the interfacial energy. The contour of the 2D projection of the function γ is represented by the orange ellipse.

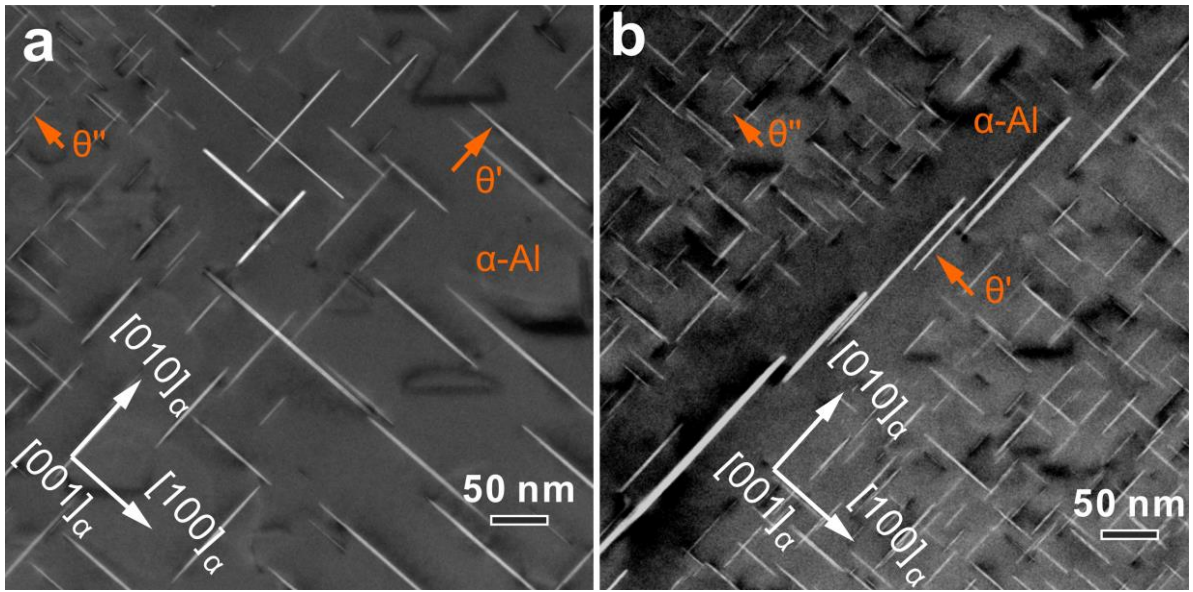

Fig. 4. HAADF-STEM images showing the distribution of θ' precipitates in an Al – 4 wt.% Cu alloy aged at 180 ºC for 66 hours. (a) Random distribution. (b) Array of parallel precipitates with the same habit plane. Viewing direction is [001]$_α$. Smaller θ" precipitates are also seen in both images.

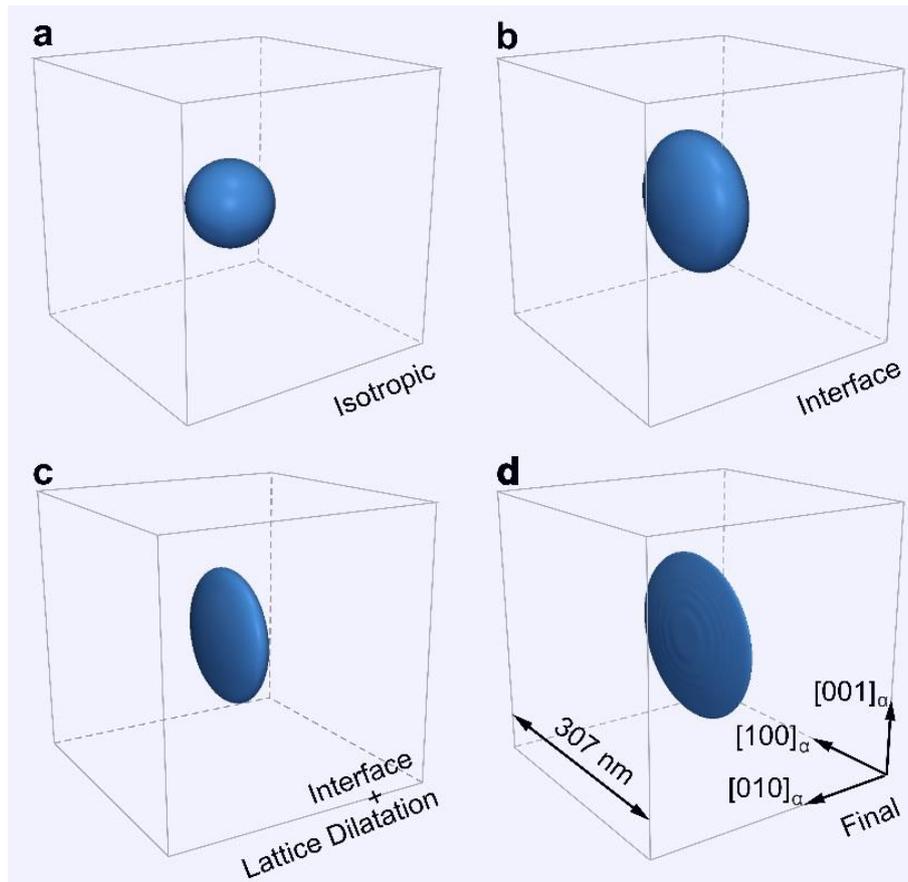

Fig. 5 Simulation results of equilibrium shape of a θ' precipitate in Al-1.74 at. % Cu alloy aged at 200 °C. The differences show the influence of the different energy contributions due to the interface and to the elastic strain associated with the transformation strain. (a) Isotropic chemical free energy. (b) Isotropic chemical free energy and anisotropic interface energy. (c) Isotropic chemical free energy and dilatational transformation strain. (d) Isotropic chemical free energy, anisotropic interface energy and dilatational and shear transformation strains.

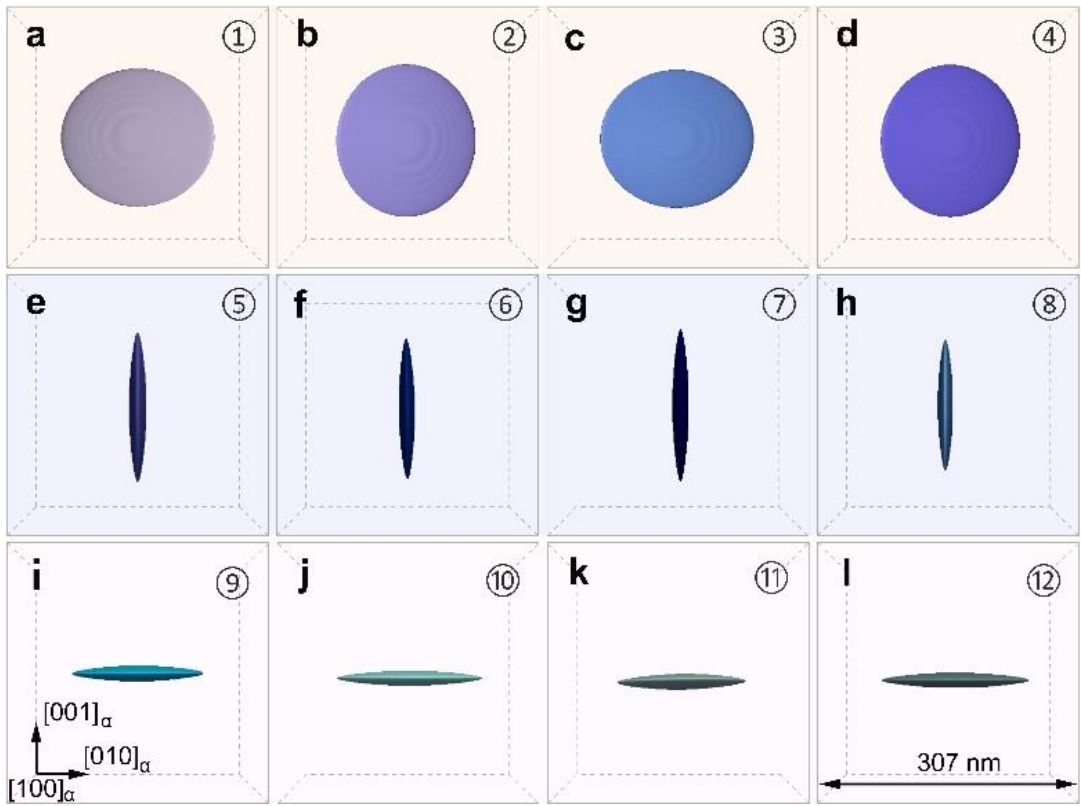

Fig. 6 Equilibrium shape of 12 deformation variants (DV) of θ' precipitates that can be grouped in 3 orientation variants OV: ①–④, ⑤–⑧, and ⑨–⑫.

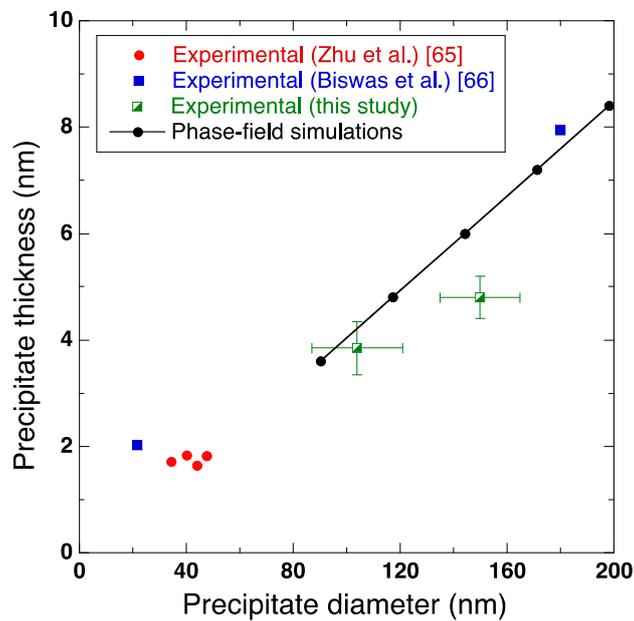

Fig. 7 Comparison between experimental results and phase-field simulations of the shape of θ' precipitates (as given by the average diameter and thickness). The experimental data were obtained in Al – 4 wt. % Cu alloys.

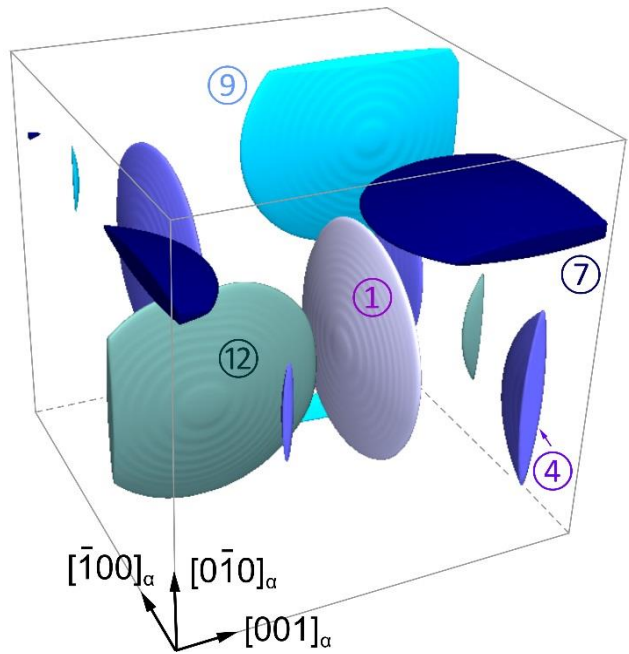

Fig. 8. Phase-field simulation of the homogeneous nucleation of θ' precipitates, leading to a random precipitate distribution. The numbers indicate the DV of each precipitate. Random noise was used for simulating the nucleation process.

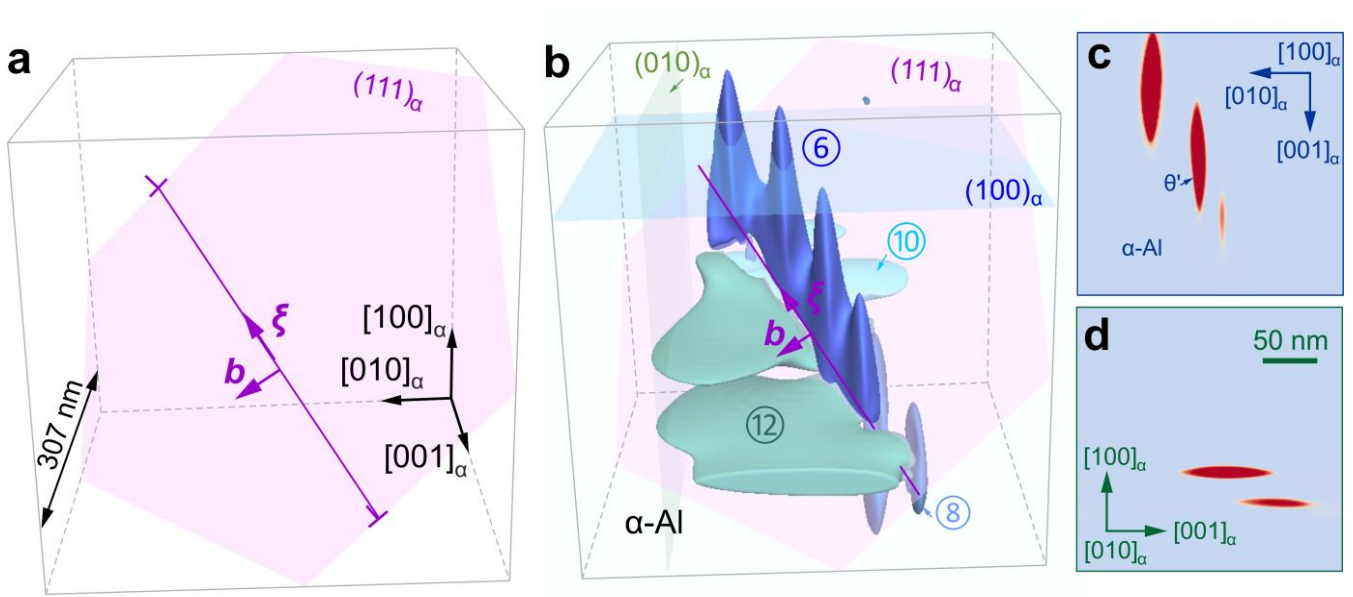

Fig. 9 (a) Edge dislocation with Burgers vector $\boldsymbol{b} = a/2\,[\bar{1}10]_\alpha$ in the $(111)_\alpha$ plane. The dislocation line is given by the vector $\boldsymbol{\xi}$. (b) Distribution of θ' precipitates along the edge dislocation in a sample aged at 200 °C at $t = 1500$. (c) Cross section of the precipitates across the $(010)_\alpha$ plane. (d) Cross section of the precipitates across $(100)_\alpha$. The cross-section planes are shown in (b).

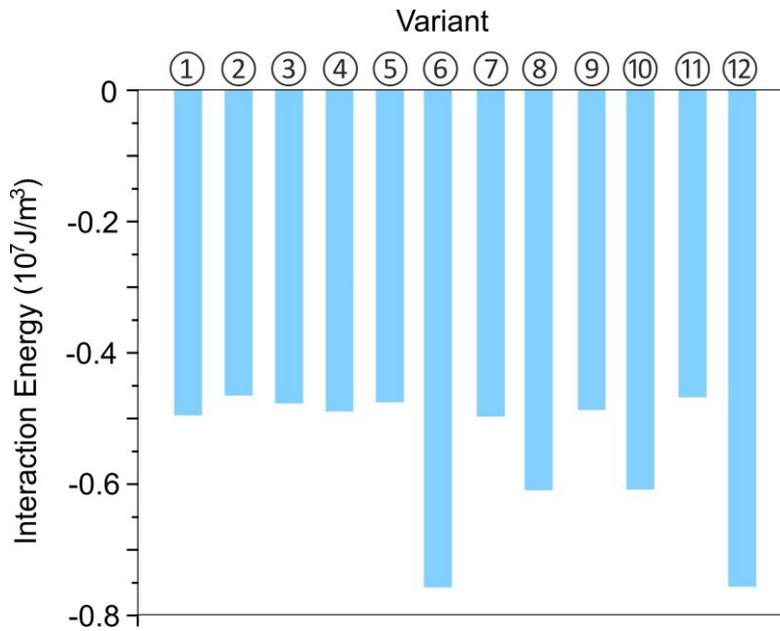

Fig. 10. Minimum interaction energies between the stress field of the pre-existing edge dislocation $\boldsymbol{b} = a/2\,[\bar{1}10]_\alpha$ in the $(111)_\alpha$ plane and the DVs of θ' precipitates.

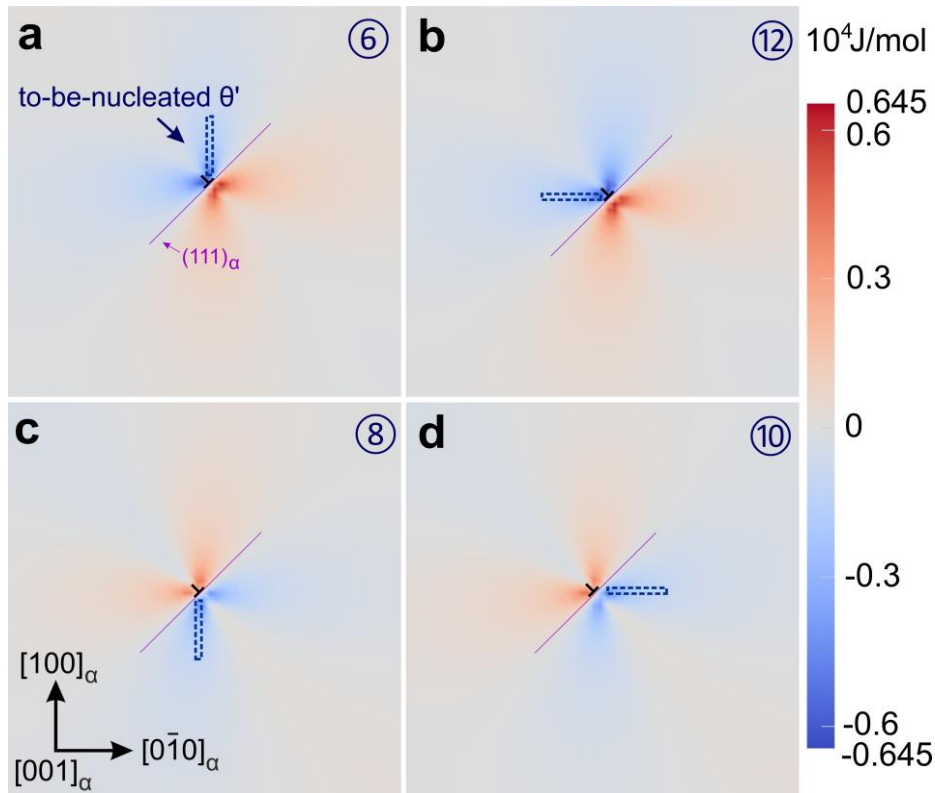

Fig. 11 Contour plot of the interaction energy between the stress field of a edge dislocation and the SFTS of different DV of the θ' precipitate in the $(001)_\alpha$ plane. (a) DV ⑥. (b) DV ⑫. (c) DV ⑧ and (d) DV ⑩. In each figure, the position of the edge dislocation is marked with the "⊥" symbol and the corresponding $(111)_\alpha$ slip plane is represented by the purple line. The shapes of the θ' variants are schematically shown by a dashed rectangle.

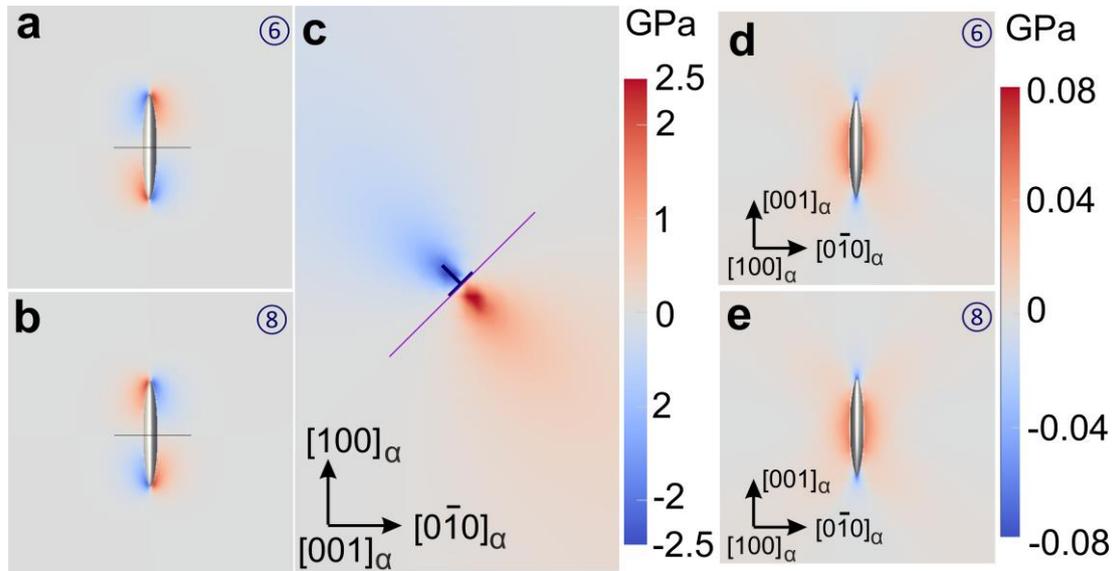

Fig. 12. Contour plots of different stress fields in the $(001)_\alpha$ plane. (a) Mean stress induced by the SFTS of precipitate belonging to DV ⑥. (b) *Idem* for DV ⑧. (c) Mean stress field of the edge dislocation. the position of the edge dislocation is marked with the "⊥" symbol and the corresponding $(111)_\alpha$ slip plane is represented by the purple line. (d) Contour plot of the mean stress in the $(100)_\alpha$ plane induced by the SFTS of precipitate belonging to DV ⑥. (e) *Idem* for DV ⑧.

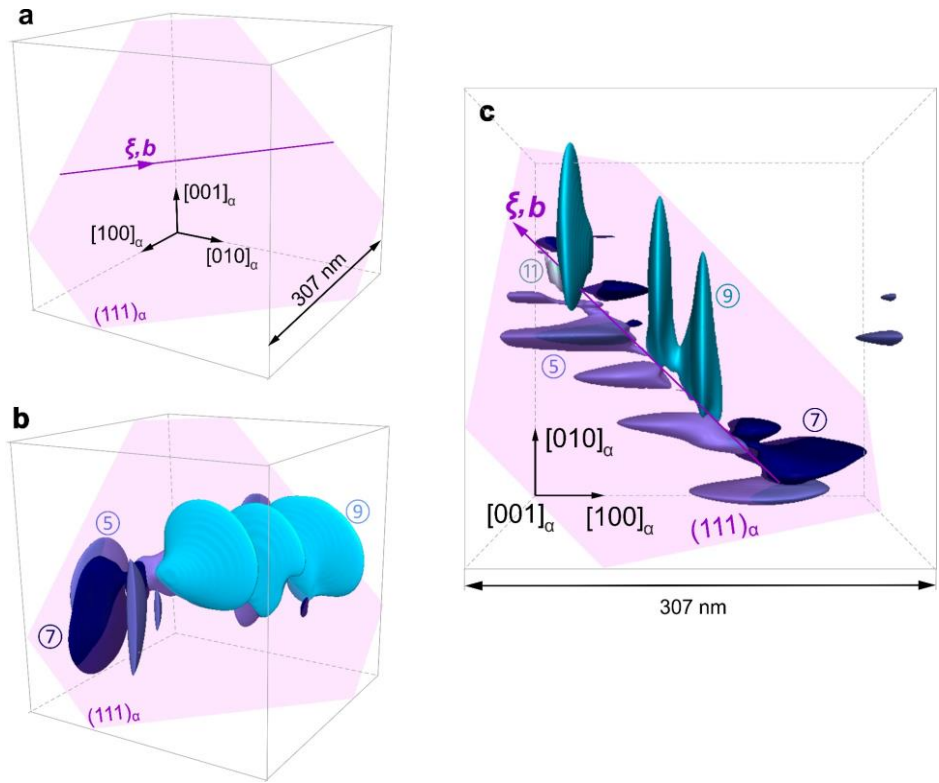

Fig. 13 (a) Screw dislocation with Burgers vector $\boldsymbol{b} = a/2\,[\bar{1}10]_\alpha$ in the $(111)_\alpha$ plane. The dislocation line is given by $\xi$. (b, c) Distribution of θ' precipitates formed under the stress field of the screw dislocation in a sample aged at 200 °C at $t = 1500$.

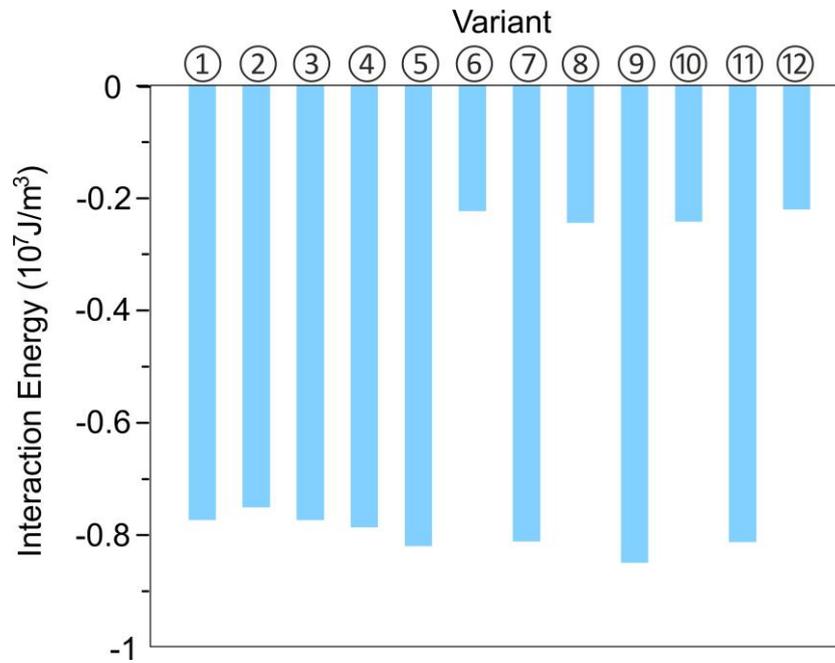

Fig. 14 Interaction energies between the stress field of the pre-existing screw dislocation $\boldsymbol{b} = a/2\,[\bar{1}10]_\alpha$ in the $(111)_\alpha$ plane and the DVs of θ' precipitates.

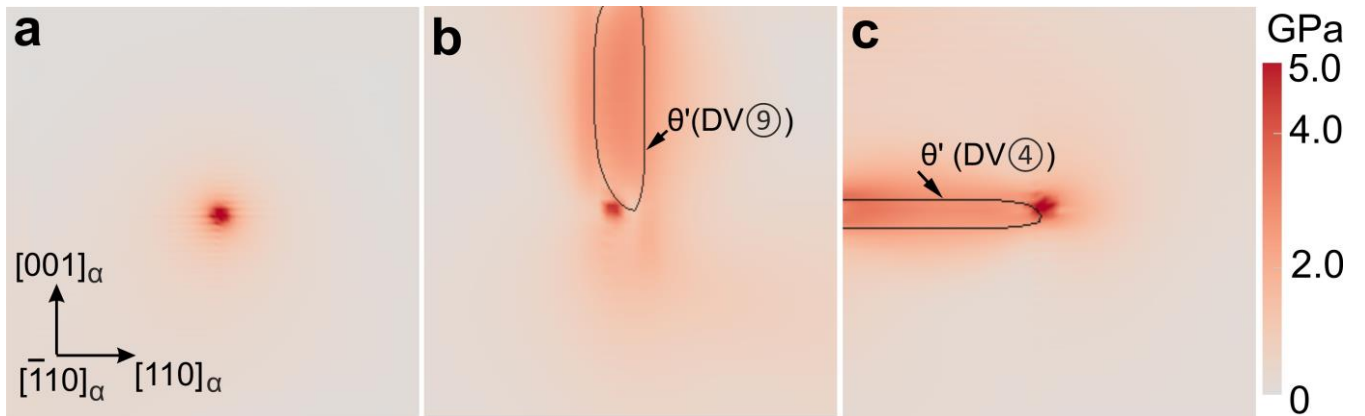

Fig. 15. (a) Contour plot of the deviatoric stress field on the $(110)_\alpha$ plane around a pre-existing screw dislocation. (b) *Idem* after adding the stress field induced by DV ⑨. (c) *Idem* after adding the stress field induced by DV ④. The black lines in (b) and (c) show the outline of each variant.

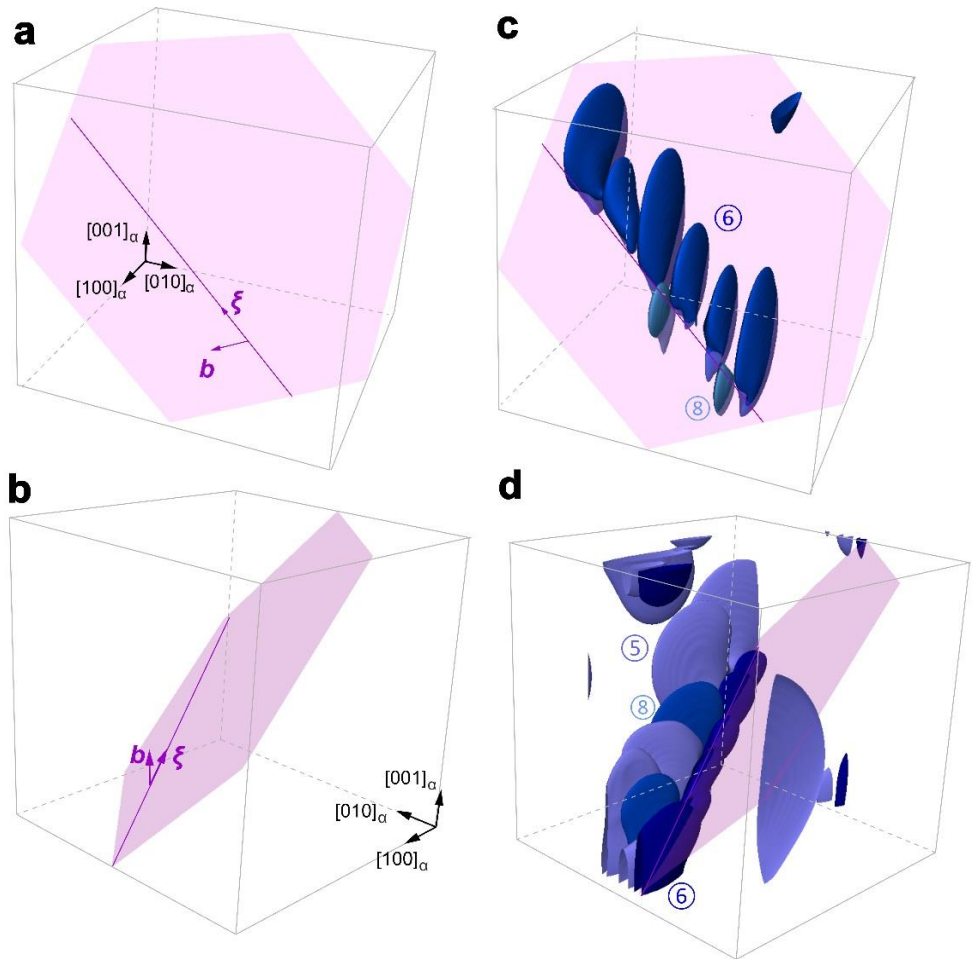

Fig. 16 (a) Mixed dislocation in the $(111)_\alpha$ plane, the angle between the Burgers vector $\boldsymbol{b} = a/2\,[\bar{1}10]_\alpha$ and the dislocation line is 60º. (b) Mixed dislocation in the $(111)_\alpha$ plane, the angle between the Burgers vector $\boldsymbol{b} = a/2\,[\bar{1}10]_\alpha$ and the dislocation line is 60º. (c-d) Distribution of θ' precipitates formed under the stress field of the mixed dislocations in (a) and (b), respectively, in a sample aged at 200 °C at $t = 1500$.

Table 1. Transformation matrix, *T*, SFTS *ε*, lattice correspondences, orientation relationships, habit planes and Orientation Variants of the 12 Deformation Variants (DV) of θ' precipitates in the α-Al matrix.

| *p* (DV) | Transformation Matrix ($T_p$) | SFTS ($\varepsilon_p$) | Lattice Correspondences | Orientation Relationship | Habit Plane | *p* (OV) |
|---|---|---|---|---|---|---|
| ① | $\begin{pmatrix} 1 & 0 & 0 \\ 0 & 1 & -0.3333 \\ 0 & 0 & 0.9384 \end{pmatrix}$ | $\begin{pmatrix} 0 & 0 & 0 \\ 0 & 0 & -0.1667 \\ 0 & -0.1667 & -0.0042 \end{pmatrix}$ | $[013]_\alpha \to [001]_{\theta'}$ $[010]_\alpha \to [010]_{\theta'}$ | | | |
| ② | $\begin{pmatrix} 1 & 0 & -0.3333 \\ 0 & 1 & 0 \\ 0 & 0 & 0.9384 \end{pmatrix}$ | $\begin{pmatrix} 0 & 0 & -0.1667 \\ 0 & 0 & 0 \\ -0.1667 & 0 & -0.0042 \end{pmatrix}$ | $[103]_\alpha \to [001]_{\theta'}$ $[100]_\alpha \to [100]_{\theta'}$ | $(001)_\alpha //(001)_{\theta'}$ $[010]_\alpha //[010]_{\theta'}$ | $(001)_\alpha$ | ❶ |
| ③ | $\begin{pmatrix} 1 & 0 & 0 \\ 0 & 1 & 0.3333 \\ 0 & 0 & 0.9384 \end{pmatrix}$ | $\begin{pmatrix} 0 & 0 & 0 \\ 0 & 0 & 0.1667 \\ 0 & 0.1667 & -0.0042 \end{pmatrix}$ | $[0\bar{1}3]_\alpha \to [001]_{\theta'}$ $[010]_\alpha \to [010]_{\theta'}$ | | | |
| ④ | $\begin{pmatrix} 1 & 0 & 0.3333 \\ 0 & 1 & 0 \\ 0 & 0 & 0.9384 \end{pmatrix}$ | $\begin{pmatrix} 0 & 0 & 0.1667 \\ 0 & 0 & 0 \\ 0.1667 & 0 & -0.0042 \end{pmatrix}$ | $[\bar{1}03]_\alpha \to [001]_{\theta'}$ $[100]_\alpha \to [100]_{\theta'}$ | | | |
| ⑤ | $\begin{pmatrix} 1 & 0 & 0 \\ 0 & 0.9384 & 0 \\ 0 & 0.3333 & 1 \end{pmatrix}$ | $\begin{pmatrix} 0 & 0 & 0 \\ 0 & -0.0042 & 0.1667 \\ 0 & 0.1667 & 0 \end{pmatrix}$ | $[03\bar{1}]_\alpha \to [001]_{\theta'}$ $[100]_\alpha \to [010]_{\theta'}$ | | | |
| ⑥ | $\begin{pmatrix} 1 & 0.3333 & 0 \\ 0 & 0.9384 & 0 \\ 0 & 0 & 1 \end{pmatrix}$ | $\begin{pmatrix} 0 & 0.1667 & 0 \\ 0.1667 & -0.0042 & 0 \\ 0 & 0 & 0 \end{pmatrix}$ | $[\bar{1}30]_\alpha \to [001]_{\theta'}$ $[001]_\alpha \to [100]_{\theta'}$ | $(010)_\alpha //(001)_{\theta'}$ $[100]_\alpha //[010]_{\theta'}$ | $(010)_\alpha$ | ❷ |
| ⑦ | $\begin{pmatrix} 1 & 0 & 0 \\ 0 & 0.9384 & 0 \\ 0 & -0.3333 & 1 \end{pmatrix}$ | $\begin{pmatrix} 0 & 0 & 0 \\ 0 & -0.0042 & -0.1667 \\ 0 & -0.1667 & 0 \end{pmatrix}$ | $[031]_\alpha \to [001]_{\theta'}$ $[100]_\alpha \to [010]_{\theta'}$ | | | |
| ⑧ | $\begin{pmatrix} 1 & -0.3333 & 0 \\ 0 & 0.9384 & 0 \\ 0 & 0 & 1 \end{pmatrix}$ | $\begin{pmatrix} 0 & -0.1667 & 0 \\ -0.1667 & -0.0042 & 0 \\ 0 & 0 & 0 \end{pmatrix}$ | $[130]_\alpha \to [001]_{\theta'}$ $[001]_\alpha \to [100]_{\theta'}$ | | | |
| ⑨ | $\begin{pmatrix} 0.9384 & 0 & 0 \\ 0 & 1 & 0 \\ 0.3333 & 0 & 1 \end{pmatrix}$ | $\begin{pmatrix} -0.0042 & 0 & 0.1667 \\ 0 & 0 & 0 \\ 0.1667 & 0 & 0 \end{pmatrix}$ | $[30\bar{1}]_\alpha \to [001]_{\theta'}$ $[010]_\alpha \to [100]_{\theta'}$ | | | |
| ⑩ | $\begin{pmatrix} 0.9384 & 0 & 0 \\ -0.3333 & 1 & 0 \\ 0 & 0 & 1 \end{pmatrix}$ | $\begin{pmatrix} -0.0042 & -0.1667 & 0 \\ -0.1667 & 0 & 0 \\ 0 & 0 & 1 \end{pmatrix}$ | $[310]_\alpha \to [001]_{\theta'}$ $[001]_\alpha \to [010]_{\theta'}$ | $(100)_\alpha //(001)_{\theta'}$ $[001]_\alpha //[010]_{\theta'}$ | $(100)_\alpha$ | ❸ |
| ⑪ | $\begin{pmatrix} 0.9384 & 0 & 0 \\ 0 & 1 & 0 \\ -0.3333 & 0 & 1 \end{pmatrix}$ | $\begin{pmatrix} -0.0042 & 0 & -0.1667 \\ 0 & 0 & 0 \\ -0.1667 & 0 & 0 \end{pmatrix}$ | $[301]_\alpha \to [001]_{\theta'}$ $[010]_\alpha \to [100]_{\theta'}$ | | | |
| ⑫ | $\begin{pmatrix} 0.9384 & 0 & 0 \\ 0.3333 & 1 & 0 \\ 0 & 0 & 1 \end{pmatrix}$ | $\begin{pmatrix} -0.0042 & 0.1667 & 0 \\ 0.1667 & 0 & 0 \\ 0 & 0 & 0 \end{pmatrix}$ | $[3\bar{1}0]_\alpha \to [001]_{\theta'}$ $[001]_\alpha \to [010]_{\theta'}$ | | | |

*p(DV)*: Deformation variant number ; *p*(OV): Orientation variant number